\newcommand{\version}{December 12, 2012}
\documentclass[a4paper,twoside,11pt,pdftex]{article}
\usepackage[bindingoffset=0.4cm,textheight=22.5cm,hdivide={2.7cm,*,2.75cm}, vdivide={*,22cm,*}]{geometry}
\usepackage{amsmath,amsfonts,amssymb}
\usepackage[pdftex]{graphicx}
\usepackage{simplewick}			% package implementing Wick-contractions
\IfFileExists{dsfont.sty}
	{\usepackage{dsfont}
         \let\mathbb=\mathds
         \newcommand{\id}{\mathds{1}}}
	{\typeout{Package dsfont.sty was not found, using alternative macros.}
         \let\mathds=\mathbb
         \newcommand{\id}{\mbox{1 \kern-.59em {\rm l}}}}

\usepackage[pdftex,bookmarksnumbered=true,breaklinks=true]{hyperref}
\usepackage[numbers,square,comma,sort&compress]{natbib}

\newcommand{\vp}{\varphi}
\newcommand{\vpb}{\bar{\varphi}}
\newcommand{\gl}{\lambda}

\newcommand{\prop}{G}
\newcommand{\IDelta}{\Delta}

\newcommand{\moyal}{(Groenewold-)Moyal}
\newcommand{\uim}{UV/IR mixing}
\newcommand{\nc}{non-commu\-ta\-tive}

\newcommand{\Tr}{\textrm{Tr}}

\newcommand{\eqnref}[1]{Eqn.~(\ref{#1})}		% for equations with preceding Eqn.
\newcommand{\secref}[1]{Section~\ref{#1}}		% for sections
\newcommand{\appref}[1]{Appendix~\ref{#1}}		% for appendix references
\newcommand{\co}[2]{\left[#1,#2\right]}					% commutator
\newcommand{\var}[2]{\frac{\d #1}{\d #2}}				% variational deriv. , p1=numerator, p2=denom.
\newcommand{\pa}{\partial}						% partial derivative sign
\newcommand{\diff}[2]{\frac{\pa #1}{\pa #2}}				% partial derivative, p1=numerator, p2=denom.
\newcommand{\ri}{{\rm i}}						% complex unit
\newcommand{\re}{{\rm e}}						% Euler e
\renewcommand{\k}{\tilde{k}}						% k tilde
\newcommand{\p}{\tilde{p}}						% p tilde
\newcommand{\q}{\tilde{q}}						% q tilde
\renewcommand{\a}{\alpha}
\renewcommand{\b}{\beta}
\newcommand{\g}{\gamma}
\renewcommand{\d}{\delta}
\newcommand{\e}{\epsilon}
\newcommand{\vare}{\varepsilon}

\renewcommand{\th}{\theta}
\renewcommand{\l}{\lambda}
\newcommand{\m}{\mu}
\newcommand{\n}{\nu}

\renewcommand{\r}{\rho}

\newcommand{\G}{\Gamma}

 \newcommand{\cK}{\mathcal{K}} \newcommand{\cL}{\mathcal{L}}
 \newcommand{\cN}{\mathcal{N}} \newcommand{\cO}{\mathcal{O}}

\newcommand{\R}{\mathds{R}}

\newcommand{\M}{\mathds{M}}

\newcommand{\inv}[1]{\frac{1}{#1}}				% inverse of something
\newcommand{\tinv}[1]{\tfrac{1}{#1}}
\newcommand{\intk}{\int\! d^4k}					% 4-dim k Fourier integral
\newcommand{\intx}{\int\! d^4x}						% 4-dim x integral
\newcommand{\nn}{\nonumber}

\newcommand{\wsq}{\widetilde{\square}}

\newcommand{\bra}[1]{\langle #1 |}
\newcommand{\ket}[1]{| #1 \rangle}

\newcommand{\phh}{\hat{\phi}}

\newtheorem{example}{Example}
\title{\begin{flushright}
       \small{UWThPh-2012-25}
       \end{flushright}
\vspace{1.5em}
\Huge On the Renormalization of Non-Commutative Field Theories}

\author{Daniel N. Blaschke\footnotemark[1]~, Thomas Garschall\footnotemark[2]~, Fran\c{c}ois Gieres\footnotemark[3]~, Franz Heindl\footnotemark[2]~,
\\Manfred Schweda\footnotemark[2]~ and Michael Wohlgenannt\footnotemark[1]\footnotemark[4]}
\date{\version}

\graphicspath{{./figures/}}
\begin{document}
\maketitle
\thispagestyle{empty}
\begin{center}
\renewcommand{\thefootnote}{\fnsymbol{footnote}}
\vspace{-0.3cm}\footnotemark[1]Faculty of Physics, University of Vienna\\Boltzmanngasse 5, A-1090 Vienna (Austria)\\[0.3cm]
\footnotemark[2]Institute for Theoretical Physics, Vienna University of Technology\\Wiedner Hauptstra\ss e 8-10, A-1040 Vienna (Austria)\\[0.3cm]
\footnotemark[3] Universit\'e de Lyon,
Universit\'e Lyon 1 and CNRS/IN2P3,\\
Institut de Physique Nucl\'eaire, Bat. P. Dirac,\\
4 rue Enrico Fermi, F - 69622 - Villeurbanne (France)\\[0.3cm]
\footnotemark[4]Austro-Ukrainian Institute for Science and Technology,\\
c/o TU Vienna, Wiedner Hauptstra\ss e 8-10, A-1040 Vienna (Austria)\\[0.5cm]
\ttfamily{E-mail: daniel.blaschke@univie.ac.at, garschall@hep.itp.tuwien.ac.at, gieres@ipnl.in2p3.fr, franz.m.heindl@gmail.com, mschweda@tph.tuwien.ac.at, michael.wohlgenannt@univie.ac.at}
\vspace{0.5cm}
\end{center}

\vspace{0.1cm}
\begin{abstract}
\noindent
This paper addresses three topics concerning the quantization
of non-commutative field theories (as defined in terms of the Moyal star product involving a constant tensor
describing the non-commutativity of coordinates in Euclidean space). To start with, we discuss the Quantum Action Principle
and provide evidence for its validity for non-commutative quantum field theories by showing that the equation of motion considered
as insertion in the generating functional  $Z^c[j]$ of connected Green functions makes sense (at least at one-loop level).
Second, we consider the generalization of the BPHZ renormalization scheme to  non-commutative field theories  and apply it to the case of a self-interacting real scalar field:
Explicit computations are performed at one-loop order and the generalization to higher loops is commented upon.
Finally, we discuss the renormalizability of various models for a self-interacting complex scalar field by using the approach of algebraic renormalization.
\end{abstract}
\vspace{1cm}
\vfill

\newpage
\thispagestyle{empty}
\tableofcontents

\newpage
\setcounter{page}{1}
%
%==============================================================================
\section{Introduction}
%==============================================================================
\label{sec:intro}

It is well known that the classical concept of space and time breaks down at very short distances
(at the order of the Planck length).
One of the ideas which have been put forward
to circumvent this problem is to generalize space-time to {\nc} spaces, the simplest one being the flat {\moyal} space -- see e.g.~\cite{Szabo:2001,Rivasseau:2007a,Blaschke:2010kw} for an introduction.
The latter is characterized by the commutation relation
\begin{align}
\co{\hat{x}^\m}{\hat{x}^\n}&=\ri\th^{\m\n}\id
\,, \label{eq:basic-comm}
\end{align}
where $\th^{\m\n}\!=\,$const., i.e. the space-time coordinates $x^{\mu}$
are promoted to operators $\hat{x}^\m$
which act on a Hilbert space and which fulfill a Heisenberg-type algebra.
The commutation relation (\ref{eq:basic-comm}) is invariant under translations of the space-time coordinates
and under the so-called reduced Lorentz transformations (or reduced orthogonal transformations in the Euclidean setting),
i.e. Lorentz transformation matrices which commute with the constant tensor $(\theta ^{\mu \nu})$, e.g. see reference~\cite{Grosse:2011es}.

Unfortunately, quantum field theories formulated on such spaces suffer from new types of infrared divergences which are related to the ultraviolet divergences of the model and
which cannot be removed by the introduction of masses. This complication which is referred to as the \emph{{\uim} problem}
implies that the familiar models like the $\phi ^4$-theory are
non-renormalizable when written on Moyal space\footnote{It is possible to remedy
this problem by considering supersymmetry, e.g. the Wess-Zumino model on Moyal space~\cite{Girotti:2000}.
However, in the present paper we restrict our attention to non-supersymmetric models.}~\cite{Minwalla:1999px,Susskind:2000}.
In recent years it could be shown that
this problem can be overcome for the  $\phi ^4$-theory in Euclidean Moyal space
by introducing specific additional terms into the action\footnote{
In this context, we should mention the so-called twisted approach to $\phi^4$-theory for which there has been some recent progress
(see for instance reference~\cite{deQueiroz:2012ti}), but which we will not discuss here.
}.
Thus, the harmonic model introduced by Grosse and Wulkenhaar~\cite{Grosse:2003,
%, Grosse:2004b,
Rivasseau:2006b}
or the $1/p^2$-model devised by
Gurau, Magnen, Rivasseau and Tanasa~\cite{Rivasseau:2008a} could be proven to be renormalizable by using the
methods of multi-scale analysis\footnote{In the Grosse-Wulkenhaar case,
the initial proof was established by its authors
by using the Wilson-Polchinski renormalization group approach in a matrix
base. A later proof~\cite{Rivasseau:2006b} relied on multi-scale analysis.}.

Several models for gauge field theories on {\nc} space have also been proposed and discussed~\cite{Grosse:2007,Wulkenhaar:2007,Blaschke:2007b,Wallet:2008a,Blaschke:2008a,Vilar:2009,Blaschke:2009e,Blaschke:2010ck},
but so far their renormalizability could not be established. In fact, the
methods considered for scalar field theories such as the multi-scale analysis cannot be applied,
or at least not without some serious complications,
since these methods break the gauge symmetry.
For commutative quantum field theories (QFTs), the approach of
algebraic renormalization~\cite{Piguet:1995,Schweda-book:1998} is a powerful tool for proving the  renormalizability of models featuring symmetries.
However, the standard formulation of this approach only applies to local field theories. Since non-commutative QFTs are non-local, the method
of algebraic renormalization cannot be used unless
it is adapted/generalized to the particular non-localities appearing in the {\nc} case.
Although this task is far from being trivial, the present paper aims to provide
some steps in this direction.

Our paper is organized as follows. In \secref{sec:QAP}, we briefly review
the Quantum Action Principle (QAP) in commutative field theory
and then we check its validity at the lowest order of perturbation theory for a
$\phi^4$-theory on
non-commutative Minkowski space
(\secref{sec:nc-scalar}).
The second topic we address is the application of
BPHZ subtraction and renormalization scheme to
field theory
on {\nc} Euclidean space
(\secref{sec:BPHZ-nc}).
The nature of non-localities in {\nc} models is analyzed in \secref{sec:locality}
before considering the approach of algebraic renormalization to
a complex scalar field theory with a rigid $U(1)$ symmetry in \secref{sec:AR-nc}.

%%%%%%%%%%%%%%%%%%%%%%%%%%%%%%%%%%%%%%%%%%%%%%%
\subsection{Brief review of models in Euclidean space}\label{briefreview}

When discussing different aspects of the renormalization of {\nc} QFTs,
it is useful to have in mind the specific models which have been considered and the results which have been established
concerning their renormalizability.
We therefore present a brief review of models while focussing on real scalar fields on Euclidean Moyal space in four dimensions. (For other dimensions and for fermionic fields, see for instance~\cite{deGoursac:2009gh} and references therein.)
The case of a complex scalar field will be discussed in \secref{sec:AR-nc}.
The perturbative quantization in {\nc} Minkowski space and the passage to  Euclidean space
is commented upon in the next subsections.

\noindent
\textbf{Na\"ive $\phi^4$-theory:} The action is given by
 \begin{align}
\label{naiveaction}
S_0[\phi] = \int d^4x \left( \inv2\, \partial^\mu \phi \,  \partial_\mu \phi + \inv2 \, m^2  \phi^2
+  \frac{\lambda}{4!} \, \phi\star\phi\star\phi\star\phi \right)\,,
\end{align}
and this model cannot be expected to be renormalizable due to the {\uim} problem.

\noindent
\textbf{Harmonic model for the $\phi ^4$-theory:}
A renormalizable (though not translation invariant) model for the $\phi ^4$-interaction
is obtained~\cite{Grosse:2003,
%Grosse:2004b,
Rivasseau:2006b}
by adding a harmonic term to the na\"ive action (\ref{naiveaction}):
 \begin{align}
\label{GWaction}
S_{GW} [\phi] = S_0[\phi] +  \int d^4x \left( \inv2\, \Omega^2  (\tilde{x}^{\mu} \phi ) \star   (\tilde{x}_{\mu} \phi ) \right)
\,.
\end{align}
Here, $\tilde{x} _{\mu} \equiv 2 \left( \theta ^{-1} \right) _{\mu \nu} x^{\nu}$ and $\Omega >0$ is a dimensionless constant.
In the particular case where $\Omega =1$, the action (\ref{GWaction}) is invariant under the so-called Langmann-Szabo duality.
We note that the harmonic term admits a geometric interpretation in terms of {\nc} scalar curvature~\cite{deGoursac:2010zb}.

\noindent
\textbf{Harmonic model for the $\phi ^3$-theory:}
The action reads
\begin{align}
\label{eq:naiveaction}
S_3 [\phi] = \int d^4 x \left(
\inv2 \, \partial^\mu \phi \,  \partial_\mu \phi
+ \inv2 \, m^2  \phi^2
+  \inv2\, \Omega^2   (\tilde{x}^{\mu} \phi ) \star   (\tilde{x}_{\mu} \phi )
+ \frac{\lambda}{3!} \, \phi \star \phi \star \phi \right)
\,,
\end{align}
and is renormalizable~\cite{Grosse:2006qv}.

\noindent
\textbf{$\phi ^4$-theory with a $1/p^2$-term:}
A translation invariant renormalizable model for the quartic self-interaction
originates from the inclusion of a non-local counterterm  which eliminates
the most singular part of the IR divergence and thereby overcomes the problematic {\uim} problem~\cite{Rivasseau:2008a}:
 \begin{align}
\label{Gurauaction}
S_{\text{trans.inv.}} [\phi] = S_0[\phi] -  \int d^4x \left( \phi \, \frac{a^2}{\Box}  \,  \phi \right)
\,.
\end{align}
Here, the parameter $a$ is assumed to have the form $a = a' / \theta$ where $a'$ represents a real dimensionless constant
and the deformation matrix $(\theta ^{\mu \nu})$
can be (and is) assumed to have the simple block-diagonal form
 \begin{align}
\label{blockdiag}
(\theta ^{\mu \nu}) = \theta \left(
\begin{array}{cccc}
0 & 1 & 0  & 0 \\
-1 & 0  & 0  & 0 \\
0 & 0  & 0  & 1 \\
0 & 0  & -1  & 0
\end{array}
\right)
\, , \qquad {\rm with} \ \; \theta \in \R
\, .
\end{align}

\noindent
\textbf{$\phi ^4$-theory on degenerate Moyal space:}
Motivated by the work~\cite{Wang:2007uq}, Grosse and Vignes-Tourneret~\cite{Grosse:2008df}
studied the  $\phi ^4$-theory on a degenerate Moyal space (i.e. a space for which some of the coordinates commute with each other).
They found that in this case the addition of the harmonic term is not sufficient to establish renormalizability
-- an extra term of the type  $\inv{\theta ^2} \, ({\rm Tr}\,  \phi)^2 $ is required: the action
  \begin{align}
\label{grossevignes}
S_{GVT} [\phi] = & \int d^2x  \int d^2y
\, \inv2 \, \phi (\vec x, \vec y \, )
\left( - \Delta + m^2 + \frac{\Omega^2}{\theta^2}  \,  \vec y ^{\; 2} \right) \phi (\vec x, \vec y \, )
\nn\\
&  + \frac{\kappa^2}{\theta^2} \,   \int d^2x  \int d^2y  \int d^2z
\, \phi (\vec x, \vec y \, ) \phi (\vec x, \vec z \, )
+ \frac{\lambda}{4!}
 \int d^4x \,  \phi\star\phi\star\phi\star\phi
\end{align}
is renormalizable.
This result is of interest in relation with the attempts to construct renormalizable models on {\nc} Minkowski space involving a time coordinate which commutes with the spatial coordinates.
More generally, terms of the type ``product of traces'' appear to be quite natural for theories on {\nc} spaces~\cite{Gayral:2006wu}.

%%%%%%%%%%%%%%%%%%%%%%%%%%%%%%%%%%%%%%%%%%%%%%%
\subsection{Models in Minkowski space}\label{sec:minkmodels}

The perturbative quantization of interacting field theories on {\nc} Minkowski space has been studied in various works
while starting from free field theory --- see for instance reference~\cite{Micu:2000} for the $\phi^4$-theory.
The Moyal star product of fields implies that the latter no longer commute
at space-like separated points, i.e. one has a violation of
microcausality. This fact is at the origin of a wealth of conceptual and technical problems (notably with unitarity), e.g. see references~\cite{Denk:2003jj}.

%%%%%%%%%%%%%%%%%%%%%%%%%%%%%%%%%%%%%%%%%%%%%%%
\subsection{Wick rotation for {\nc} field theories}\label{sec:nc-wick-rot}

The passage from a field theory  on {\nc} Euclidean space to one on {\nc} Minkowski space
represents a subtle issue since the Wick rotation (as defined in usual quantum field theory) relies on the properties of locality and covariance.
For an invertible deformation matrix $(\theta ^{\mu \nu})$, some partial results have recently been established~\cite{Fischer:2008dq},
but so far no final conclusion has been obtained.
For the case where time remains commutative (i.e. degenerate Moyal space), the algebraic
approach to quantum field theory~\cite{Haag:1992hx} was considered quite recently by the authors of reference~\cite{Grosse:2011es}
to obtain a condition allowing for an analytic continuation of field theory from Euclidean to Minkowskian Moyal space.
The results are based on the assumption of the so-called \emph{time zero condition} which appears to be a strong constraint restricting the class of models which may be considered.
The investigation of specific models represents an ongoing discussion~\cite{Grosse:2012my}.

%
%==============================================================================
\section{QAP for commutative QFTs}
%==============================================================================
\label{sec:QAP}

The Quantum Action Principle (QAP),
i.e. the renormalized version of Schwinger's action principle, plays an important role in the
 study of field theories which are invariant under transformations that depend non-linearly on the fields,
e.g. non-Abelian gauge field theories. These symmetries are expressed by Ward identities which must
be satisfied by the Green functions of the theory.
The situation is well understood for QFTs on commutative space and details concerning power counting renormalizable QFTs may
for example be found in the references~\cite{Piguet:1995,  Schweda-book:1998, haeussling}.
Quite generally the QAP describes the result of the insertion of a composite operator into Green functions
in the renormalized theory.
There are three versions of the QAP corresponding to the different types of insertions
to be considered: equations of motion (e.o.m.),
local field polynomials and derivatives with respect to parameters.
In the following we briefly recall these three versions by considering the example of the $\phi^4$-theory,
i.e. the power counting renormalizable scalar field theory on commutative Minkowski space-time $\M^4$
defined by the classical action\footnote{We consider the signature $(+,-,-,-)$ for the Minkowski metric and we use the natural system of units ($\hbar \equiv 1 \equiv c$).}
\begin{align}
S[\phi]&=  S_0[\phi] + S_{\rm int} [ \phi] \equiv \intx \, (- \inv2 \, \phi \, \cK_x \phi ) + \intx \, ( - \frac{\gl}{4!} \, \phi^4 )
\, ,
\end{align}
where
\begin{align}
 \cK_x=\square_x+m^2
 \,. \label{eq:def-Kx}
\end{align}
The fundamental quantities in quantum theory are the Green functions,
in particular the connected Green functions: the latter are collected in a generating functional
$Z^c[j]=-\ri\ln Z[j]$  from which they can be extracted by differentiating with respect to the external source $j$,
\begin{align}
 \frac{\d^nZ^c[j]}{\d j(x_1)\ldots\d j(x_n)} \, \Big|_{j=0}&=\ri^{n-1}\bra{0}T\phh(x_1)\ldots\phh(x_n)\ket{0}_c
\,,
\end{align}
where the $\phh(x_i)$ are Heisenberg field operators, see for instance reference~\cite{Das:2006}.

\subsection*{Insertion of e.o.m.}
%%%%%%%%%%%%%%%%%%%%%%%%%%%%%%%%%

Starting from the generating functional $Z[j]=\exp(\ri Z^c[j])$ of Green functions, a formal calculation leads to an equation representing the e.o.m. of the theory expressed in functional form:
\begin{align}
 \var{S[\phi]}{\phi(x)}\Big|_{\phi\to\frac1\ri\var{}{j}}Z^c[j]+j(x)&=0
\,. \label{eq:class-funct-eom}
\end{align}
By using any of the known renormalization schemes (BPHZ, dimensional
or analytic regularization, \ldots),
one can prove that an equation of the form \eqref{eq:class-funct-eom} holds in the renormalized theory, i.e. the statement which is referred to as the first version of the QAP:
\begin{align}
O(x)\cdot Z^c[j]+j(x)&=0
\,. \label{eq:qap}
\end{align}
Here, $ O(x)\cdot Z^c[j]$ denotes the insertion of the local composite operator $\hat{O}(x)$
into the renormalized connected Green functions, i.e.
\begin{align}
\bra{0}T\phh(x_1)\ldots\phh(x_n)\ket{0}_c \leadsto \bra{0}T \hat{O}(x) \phh(x_1)\ldots\phh(x_n)\ket{0}_c
\,,
\end{align}
and at the lowest order in the $\hbar$-expansion $\hat{O}(x)$ is the local classical field polynomial
$\delta S/ \delta \phi(x)$ (whose vanishing is tantamount to the classical e.o.m.):
\begin{align}
\hat{O}(x) = \widehat{\var{S[\phi]}{\phi(x)}} +\cO(\hbar ) =-\cK_x\phh (x)- \frac{\gl}{3!} \, \phh^3  (x)
+\cO(\hbar )
\,. \label{eq:qap-eom-general}
\end{align}
Furthermore, the dimension of $\hat{O}(x)$ is bounded by $4 - {\rm dim} \, \phh$.

For the \emph{free theory} (i.e. for $\gl =0$), \eqnref{eq:qap} is nothing but the free
e.o.m. in the presence of the source $j$.

\subsection*{Insertion of field polynomials}

The second version of the QAP which can be proven, consists of the insertion of $Q(x) \frac{\delta S [\phi]}{\d \phi (x)} +{\cal O}(\hbar )$ where $Q(x)$ is a classical field polynomial.

\subsection*{Insertion of derivatives with respect to parameters}

The third version comprises of the insertion of $\frac{\partial S }{\partial \lambda}+{\cal O}(\hbar )$ where $\lambda$ denotes a parameter appearing in $S$.

\subsection*{Some consequences}
%%%%%%%%%%%%%%%%%%%%%%%%%%%%%%%%

The first version of the QAP as given by \eqnref{eq:qap} states that the functional $F[j]$
appearing on the lhs of this equation identically vanishes or equivalently, that all of its Taylor
coefficients vanish, i.e.
\[
 \frac{\d^n F[j]}{\d j(x_1)\ldots\d j(x_n)} \, \Big|_{j=0} =0 \, , {\qquad} \ \; {\rm for} \ n=0,1, \dots
\]
Each of the latter identities can be viewed as a perturbative expansion in the coupling constant $\gl$.
In the following,
we will explicitly derive the relations that one obtains for $n=1$ at the lowest orders in $\gl$.
These identities are ensured to hold as mere consequences of the QAP whose validity is well established.
In \secref{sec:nc-scalar}, we will then verify  explicitly the validity of the analogous consequences
for {\nc} $\phi^4$-theory at next-to-leading order in perturbation theory, thereby providing evidence
for the validity of the QAP in the {\nc} setting.

By differentiating relation (\ref{eq:qap}) with respect to $j(y)$ at $j=0$ and using $\frac{\delta j(x)}{\delta j(y)} = \delta^{(4)} (x-y)$, one concludes that
for the \emph{free scalar field theory}
\begin{align}
\ri\d^{(4)}(x-y) =
\bra{0}T\hat{O}  (x) \phh(y)\ket{0} _c
&=-\cK_x\bra{0}T\phh(x)\phh(y)\ket{0}
\,,
\label{freetheory}
\end{align}
i.e. the defining equation for the propagator whose solution reads
\begin{align}
\label{defscalprop}
\bra{0}T\phh(x)\phh(y)\ket{0} \equiv
\ri
\prop(x-y)  =\lim\limits_{\e \to0^+}\int\!\frac{d^4k}{(2\pi)^4} \re^{-\ri k(x-y)}\frac{\ri}{k^2-m^2+\ri \e}
\, .
\end{align}
For the \emph{interacting theory}, differentiation of relation (\ref{eq:qap}) with respect to $j(y)$ at $j=0$ yields
\begin{align}
\bra{0}T\left[ \cK_x\phh(x)\phh(y)+\frac{\gl}{3!}\phh^3(x)\phh(y)\right]
\ket{0}_c
+\ri\d^{(4)}(x-y)=0
\,.
\label{eq:qap-eom-phi4-example}
\end{align}
We may then apply the Gell-Mann-Low formula\footnote{Note, that the lhs involves Heisenberg field operators, whereas the rhs concerns asymptotic free fields as indicated by the subscript $0$. }
\begin{align}
\bra{0}T\phh(x_1)\ldots\phh(x_n)\ket{0}_c&=\frac{\bra{0}T\phh(x_1)\ldots\phh(x_n)\exp\left(\ri\int\!d^4z \,  \cL_{\textrm{int}}\right)\ket{0}_0}{\bra{0}T\exp\left(\ri\int\!d^4z \, \cL_{\textrm{int}}\right)\ket{0}_0}
\,,\label{eq:gellmannlow}
\end{align}
and make use of Wick contractions for free asymptotic fields:
\begin{align}
\label{wick-contraction}
T ( \phh(x) \phh(y) )
& =\; :\phh(x) \phh(y): + \,
\contraction{}{\phh(x)}{}{\phh(y)} \phh(x) \phh(y)
\,,
\nn\\
{\rm with} \qquad \contraction{}{\phh(x)}{}{\phh(y)} \phh(x) \phh(y)
& \equiv \bra{0}  T( \phh(x) \phh(y) ) \ket{0}_0
\, \id
= \ri \prop (x-y)
\, \id
\,.
\end{align}
To first order in the coupling constant, expression (\ref{eq:qap-eom-phi4-example}) then reads
\begin{align}
\bra{0}T\left[ \cK_x\phh(x) \Big( {\id} -\frac {\ri \gl}{4!} \int d^4z \, \phh^4(z) \Big) \phh(y)+\frac{\gl}{3!} \, \phh^3(x)\phh(y)\right]
\ket{0}_{0,c}
+ \ri\d^{(4)}(x-y)=0
\,, \label{eq:qap-eom-phi4-example-exp}
\end{align}
where the subscript $0,c$ means that we only consider the connected parts (the non-connected contributions being compensated by the denominator
in (\ref{eq:gellmannlow})).
By  using Wick contractions expression (\ref{eq:qap-eom-phi4-example-exp})
becomes
\begin{align}
&
\cK_x \contraction{}{\phh(x)}{}{\phh(y)}\phh(x)\phh(y) -\ri \cK_x \frac {\gl}{2} \int d^4z
\contraction{}{\phh(x)}{}{\phh(z)}\phh(x)\phh(z) \contraction{}{\phh(z)}{}{\phh(z)}\phh(z)\phh(z)
\contraction{}{\phh(z)}{}{\phh(y)}\phh(z)\phh(y)\nn\\
& +
 \frac{\gl}{2} \, \contraction{}{\phh(x)}{}{\phh(x)}\phh(x)\phh(x) \contraction{}{\phh(x)}{}{\phh(y)}\phh(x)\phh(y)
+\ri \d^{(4)}(x-y) = 0
\,.
\end{align}
At order $0$ in $\gl$, we recover relation (\ref{freetheory}). At first order in $\gl$, we obtain the result
\begin{align}
\cK_x \int\!\! d^4z\, \prop(x-z) \prop(0) \prop(z-y) + \prop(0) \prop(x-y) = 0
\,,
\end{align}
which may be interpreted graphically by:
\begin{align*}
\includegraphics[scale=0.8]{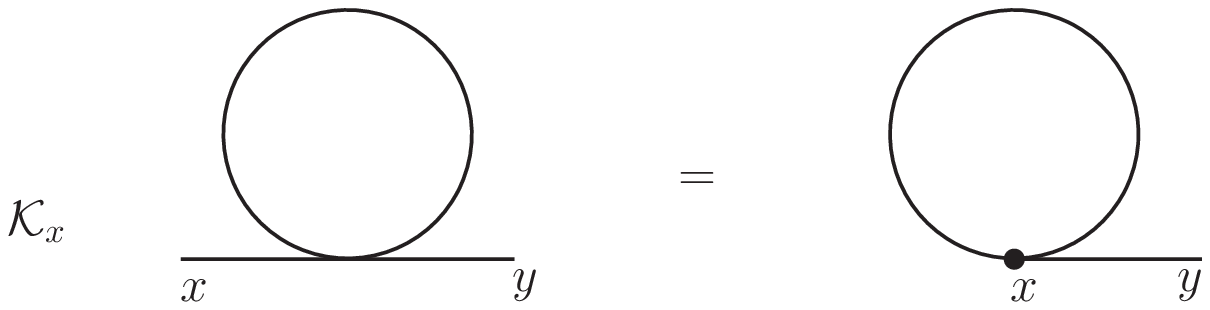}
\end{align*}

At order $\gl ^2$, the expansion (\ref{eq:qap-eom-phi4-example}) yields
\begin{align}
& \bra{0} T \bigg[\cK_x\phh(x) \frac 12 (\frac{-\ri \gl}{4!})^2 \int\! d^4z_1 \int\! d^4z_2\, \phh ^4 (z_1) \phh ^4 (z_2) \phh(y) \nn\\
& \hspace{2cm}
+ \frac{\gl}{3!} \, \phh ^3 (x) (\frac{-\ri \gl}{4!}) \int\! d^4z\, \phh ^4 (z) \phh(y) \bigg]
\ket{0}_{0,c}
= 0\,.
\end{align}
This relation as well as those obtained at higher order in $\gl$ can be explicitly worked out along the lines indicated above.

\paragraph{Non-linear field variations.}
%%%%%%%%%%%%%%%%%%%%%%%%%%%%%%%
Now let us consider field variations which we denote by $\phi\to p(\phi(x))=:P(x)$.
Eqn.~\eqref{eq:qap} is then replaced by~\cite{Piguet:1995}
\begin{align}
\left(p(\phi)\var{S[\phi]}{\phi(x)}\right)\cdot
Z^c[j,\rho ] +j(x)\var{Z^c[j,\rho ] }{\r(x)}=0
\,,
\end{align}
where $\r$ denotes an external source which is coupled linearly to $p(\phi)$.
Varying twice with respect to $j$ and setting the sources to zero afterwards, leads to the
following  relation for graphs with two external legs~\cite{Balasin-diss,Breitenlohner:1977hr}:
\begin{align}
&\bra{0}T\intx \hat{P}(x)\widehat{\var{S[\phi]}{\phi(x)}}\phh(x_1)\phh(x_2)\ket{0}_c+\bra{0}T \hat{P}(x_1)\phh(x_2)\ket{0}_c+\bra{0}T \hat{P}(x_2)\phh(x_1)\ket{0}_c=0
\,.
\label{eq:qap-nonlinear-2point}
\end{align}

\paragraph{Euclidean case.}
%%%%%%%%%%%%%%%%%%%%%%%%%%%%%%%
Before closing this section, we should mention that the QAP is also valid for field theories on Euclidean space (see e.g. reference~\cite{Piguet:1995}), the obvious difference with the Minkowskian case being factors of $\pm\ri$.
More specifically, in Euclidean
space the generating functional for the connected Green functions is given by
\begin{align}
Z^c[j]&=-\ln Z[j]
 \,,
\end{align}
and the Gell-Mann-Low formula may be expressed as
\begin{align}
 Z[j]=\cN \, \re^{-S_{\text{int}}\left[-\var{}{j}\right]}Z_{\text{free}}[j]
 \, .
\end{align}

%==============================================================================
\section{QAP for {\nc} scalar quantum field theory}
%==============================================================================
\label{sec:nc-scalar}

We will now ``switch on'' the non-commutativity, i.e. consider the non-vanishing commutator of \eqnref{eq:basic-comm}.
In order to simplify the calculations,
we employ the Weyl quantization map $\hat{\mathcal{W}}$ which enables us to
implement the non-commutativity of space in terms of a deformed product
for the functions defined on this space, namely
the {\moyal} star product (see e.g. reference~\cite{Szabo:2001} for an introduction):
\begin{align}
(f\star g)(x):=\hat{\mathcal{W}}^{-1}\!\left[\hat{\mathcal{W}}[f]\hat{\mathcal{W}}[g]\right](x)
=
\left(
e^{\frac{\ri}{2}\th^{\m\n}\pa^x_\m\pa^y_\n}f(x)g(y) \right)\! \Big|_{x=y}
\,.
\end{align}
Since
\begin{align}
\Tr\left(\hat{\mathcal{W}}[f_1]\cdots\hat{\mathcal{W}}[f_n]\right)=\intx\, (f_1\star\cdots\star f_n)(x)
\,, \label{eq:cyclic-property}
\end{align}
the integral of a star product of functions shares the property of the trace of being invariant under cyclic permutations.

For the discussion of the na\"ive $\phi^4$-theory on {\nc} Minkowski space, we will follow the conventions
of Micu and Sheikh-Jabbari~\cite{Micu:2000},
and consequently we choose $\theta^{0i}=0$ in order to avoid difficulties
with the time ordering~\cite{Gomis:2000}. %,Heslop:2004xj}.
The action for the na\"ive $\phi^4$-theory reads
\begin{align}
\label{naivephei}
S[\phi, \gl ] = \int d^4x \left[ \frac 12 ( \partial^\mu \phi \star \partial_\mu \phi - m^2 \phi\star \phi )
- \frac \gl {4!} \, \phi\star\phi\star\phi\star\phi \right]
\,,
\end{align}
and the property \eqref{eq:cyclic-property} of the star product allows us always to drop one star.
The Feynman rules for this model are given by
\begin{align}
\raisebox{-1.5pt}[1pt][0pt]{\includegraphics[scale=0.8, trim=0 0 0 10,clip=true]{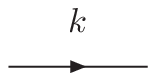}}&= \; \prop(k)=
\inv{k^2-m^2+\ri\vare}\,,
\nn\\
\hspace*{-12pt}
\raisebox{-30pt}[24pt][0pt]{\includegraphics[scale=0.8]{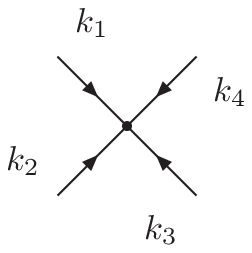}} &=  \ri V(k_1,k_2,k_3,k_4)
\nonumber \\
&=  \frac{\ri\l}{3}(2\pi)^4 \delta^{(4)}
\left(k_1+k_2+k_3+k_4\right) \Bigg[ \cos \left( \frac{1}{2} \,
k_1\k_2 \right)\cos\left(\frac{1}{2} \,k_3\k_4 \right)
\nonumber\\
& \quad +\cos\left(\frac{1}{2} \, k_1\k_3\right)\cos\left(\frac{1}{2}
\, k_2\k_4\right)+\cos\left(\frac{1}{2} \, k_1\k_4
\right)\cos\left(\frac{1}{2} \, k_2\k_3\right)\Bigg]
, \label{eq:feynmanrules-naivemodel}
\end{align}
where we have introduced the notation
\begin{align}
 \k^\m:=\th^{\m\n}k_\n
\,.
\end{align}
In the following considerations concerning QFT, we will omit the hats on the Heisenberg field operators
in order to simplify the notation.

\subsection{Verifying the QAP for the e.o.m.}
%%%%%%%%%%%%%%%%%%%%%%%%%%%%

As indicated in the previous section, we would like to provide evidence for the validity of the QAP in {\nc} field theory.
In order to do so, we assume that the latter has the standard form,
i.e. $F[j]=0$ where $F[j]$ is given by the lhs of \eqnref{eq:qap},
hence $\left. \frac{\delta F}{\delta j(x)} \right|_{j=0} =0$. In the sequel, we explicitly verify this relation
to lowest order in perturbation theory for the  na\"ive $\phi^4$-theory
in Minkowski space. A complete proof of the QAP (which is beyond the scope of the present work)
requires both the consideration of all orders of the Taylor series expansion and
a focus on renormalizable models
(in Euclidean space).

The e.o.m. associated to the action (\ref{naivephei}) is determined by
\begin{align}
O(x) \equiv \frac{\delta S[\phi, \gl ]}{\delta\phi(x)} = - \cK_x\phi(x) - \frac \gl {3!} \, \phi ^{\star 3}(x)
\, ,
\qquad {\rm with} \ \
\phi ^{\star 3} \equiv \phi\star\phi\star\phi
\,.
\label{eq:qap-eom}
\end{align}
The QAP for {\nc} quantum field theory is formulated as
\begin{align}
O(x) \cdot Z^c[j] + j(x) = 0\,.
\end{align}
Differentiation with respect to $j(y)$ at $j=0$ yields
\begin{align}
\label{eq:nc-qap-cons}
\bra{0} T \bigg[  \cK_x\phi(x) \phi(y) + \frac{\gl}{3!} \, \phi ^{\star 3} (x) \phi(y)
\bigg] \ket{0}_c + \ri\delta^{(4)} (x-y) = 0\,.
\end{align}
In order to evaluate this expression
 perturbatively using the Gell-Mann--Low formula\footnote{At this point we assume
 that the Gell-Mann--Low formula holds in the  {\nc} setting.
The computation carried out in the remainder of this section indirectly verifies this assumption up to order $\gl$.}
one needs to consider the Wick contractions in the {\nc} setting.
For convenience, we use their momentum space representation as following from
\eqnref{wick-contraction} and \eqnref{defscalprop}:
\begin{align}
 \contraction{}{\phi(x)}{}{\phi(y)}\phi(x)\phi(y)
& = \lim\limits_{\e \to0^+}\int \frac{d^4k}{(2\pi)^4} \, \re^{-\ri k(x-y)} \, \frac \ri{k^2-m^2+\ri\epsilon} \nn\\
& \equiv
 \int \frac{d^4k_1}{(2\pi)^4} \int \frac{d^4k_2}{(2\pi)^4} \,
 \re^{- \ri k_1 x - \ri k_2 y}
\, \contraction{}{\tilde \phi(k_1)}{}{\tilde \phi(k_2)}\tilde \phi(k_1)\tilde \phi(k_2)
\,.
\end{align}
This implies the identification
\begin{align}
\contraction{}{\tilde \phi(k_1)}{}{\tilde \phi(k_2)}\tilde \phi(k_1)\tilde \phi(k_2) =
\frac {(2\pi)^4 \ri}{k_1^2-m^2+\ri\epsilon} \; \delta^{(4)}(k_1+k_2)\,.
\end{align}
Let us now return to \eqnref{eq:nc-qap-cons}. As in the commutative case (see Eqn.(\ref{eq:qap-eom-phi4-example-exp})), we obtain up to first order in the coupling constant $\gl$
\begin{align}
\bra{0}T\left[\cK_x\phi(x) \Big( \id -\frac {\ri \gl}{4!} \int d^4z \, \phi ^{\star 4} (z) \Big) \phi(y)
+ \frac{\gl}{3!} \, \phi ^{\star 3} (x) \phi(y) \right]
\ket{0}_{0,c}
+ \ri\d^{(4)}(x-y)=0
\,. \label{eq:qap-eom-phi4-nc}
\end{align}
At the order zero in $\gl$, this relation represents the defining relation for the propagator.
Let us scrutinize the terms in Eqn. (\ref{eq:qap-eom-phi4-nc}) which are of order one in $\gl$:
the first of these terms reads as follows in momentum space
\begin{align}
\nonumber
& \bra{0}T \, \cK_x ( \frac{-\ri \gl}{4!} ) \int \frac{d^4 q_1}{(2\pi)^4} \int \frac{d^4 q_2}{(2\pi)^4}
\left( \prod_{j=1}^4 \int \frac{d^4 k_j}{(2\pi)^4} \right)
\tilde \phi(q_1) \tilde \phi(k_1) \tilde \phi(k_2) \tilde \phi(k_3) \tilde \phi(k_4) \tilde \phi(q_2)\\
& \hspace{.5cm} \times
\frac 13 \left[
\cos\! \left(\!\frac{k_1\k_2}2\!\right) \cos\! \left(\frac{k_3\k_4}2\!\right) + \cos\! \left(\!\frac{k_1\k_3}2\!\right) \cos\! \left(\!\frac{k_2\k_4}2\!\right)
+ \cos\! \left(\!\frac{k_1\k_4}2\!\right) \cos\! \left(\!\frac{k_2\k_3}2\!\right)
\right] \nn\\
& \hspace{.5cm} \times
\re^{- \ri q_1 x - \ri q_2 y}
\, (2\pi)^4 \, \delta^{(4)} \left( \sum_{i=1}^4 k_i \right)
\ket{0}_{0,c}
\,.
\end{align}
We can now employ Wick contractions in momentum space and obtain
\begin{align}
&
 \frac{-\ri \gl}{6} \,
 \cK_x  \iint \frac{d^4 q_1}{(2\pi)^4} \frac{d^4 q_2}{(2\pi)^4}
\left( \prod_{j=1}^4 \int d^4 k_j \right)
\re^{- \ri q_1 x -  \ri q_2 y}
\, \delta^{(4)} \left( \sum_{i=1}^4 k_i \right) \\
\nonumber
& \hspace{.5cm} \times
\frac{\ri \delta^{(4)}( q_1+k_1 )}{q_1^2 - m^2 + \ri \epsilon}
\; \frac{\ri \delta^{(4)}( k_2+k_3 )}{k_2^2 - m^2 + \ri \epsilon}\,
\; \frac{\ri \delta^{(4)}( k_4+q_2 )}{q_2^2 - m^2 + \ri \epsilon}
\\
\nonumber
& \hspace{.5cm} \times
\left[
\cos\! \left(\! \frac{k_1\k_2}2\!\right) \cos\! \left(\! \frac{k_3\k_4}2\!\right) + \cos\! \left(\! \frac{k_1\k_3}2\!\right) \cos\! \left(\! \frac{k_2\k_4}2\!\right)
+ \cos\! \left(\! \frac{k_1\k_4}2\!\right) \cos\! \left(\! \frac{k_2\k_3}2\!\right)
\right]
 \,,
\end{align}
and, after some further simplifications,
\begin{align}
&\quad
 \frac {-\ri \gl}{6} \iint \frac{d^4 q_1}{(2\pi)^4} \frac{ d^4 q_2}{(2\pi)^4} \, \frac{\delta^{(4)} (q_1+q_2)}{q_2^2 - m^2 + \ri \epsilon}
\, \re^{- \ri q_1 x - \ri q_2 y} \int d^4k_2 \, \frac{ \ri}{k_2^2 - m^2 + \ri \epsilon} \nn\\
\nonumber
& \hspace{.5cm} \times
\left[
\cos\! \left(\! \frac{q_1\k_2}2\!\right) \cos\! \left(\! \frac{k_2\q_2}2\!\right) + \cos\! \left(\! \frac{q_1\k_2}2\!\right) \cos\! \left(\! \frac{k_2\q_2}2\!\right)
+ \cos\! \left(\! \frac{q_1\q_2}2\!\right)
\right] \\
& =
\frac {- \gl}{6} \int \frac{d^4 q_1}{(2\pi)^4} \, \frac{\ri}{q_1^2 - m^2 + \ri \epsilon}
\, \re^{- \ri q_1 (x-y)} \int \frac{d^4 k_2}{(2\pi)^4}
 \, \frac{\ri}{k_2^2 - m^2 + \ri \epsilon} \left[ \cos (q_1\k_2) + 2 \right]
 \,.
 \label{eq:term-1}
\end{align}
This is the well-known tadpole contribution with planar and  non-planar parts, respectively.

To conclude, we consider the second term of order $\gl$ in Eqn.~\eqref{eq:qap-eom-phi4-nc}. In momentum space, we have
\begin{align}
\phi ^{\star 3} (x) =
\iiint \frac{d^4k_1d^4k_2d^4k_3}{(2\pi)^{12}}\,
\re^{- \ri(k_1+k_2+k_3)x} \tilde\phi(k_1)
\tilde \phi(k_2) \tilde \phi(k_3) \re^{-\frac\ri 2 k_1\k_2 - \frac\ri 2 k_1\k_3 - \frac\ri 2 k_2\k_3}\,,
\end{align}
hence we obtain
\begin{align}
& \frac{\gl}{3!} \, \bra{0} T \phi^{\star 3}(x) \phi(y) \ket{0}_{0,c}
=
 \frac{\gl}6 \int \frac{d^4 q}{(2\pi)^4} \, \frac \ri{ q^2 - m^2 + \ri\epsilon } \,  \re^{- \ri q(x-y)}
 \nn\\
& \hspace{.5cm}
\times \int \frac{d^4 k}{(2\pi)^4} \, \frac \ri{ k^2 - m^2 + \ri\epsilon } \, \left[ \cos (q\k) + 2 \right] \,.
\end{align}
This expression cancels \eqref{eq:term-1}, and thus we have demonstrated the validity of the
QAP for the e.o.m. to first order.

\subsection{Verifying the QAP for a non-linear field variation}
%%%%%%%%%%%%%%%%%%%%%%%%%%%%
Here we study the {\nc} analog of \eqnref{eq:qap-nonlinear-2point} up to one-loop level for the $\phi^{\star4}$-theory
by considering  the example
\begin{align}
p (\phi)
=\phi^{\star3}
\,.
\end{align}
Using the Gell-Mann--Low formula \eqref{eq:gellmannlow} and the lines of arguments presented
in the previous subsection (cf. Eqns. \eqref{eq:qap-nonlinear-2point}, \eqref{eq:qap-eom}), one arrives at the relation
\begin{align}
0&=\bra{0}T\intx \, \phi^{\star3}(x)\star\left[\cK_x\phi(x)\Big( \id -\frac {\ri \gl}{4!} \int d^4z \, \phi ^{\star 4}  (z) \Big)
+\frac{\gl}{3!} \, \phi^{\star3}(x)\right]\phi(x_1)\phi(x_2)\ket{0}_c \nn\\
&\quad - \bra{0}T\phi^{\star3}(x_1)\phi(x_2)\Big( \id -\frac {\ri \gl}{4!} \int d^4z \, \phi ^{\star 4} (z) \Big)\ket{0}_c \nn\\
&\quad - \bra{0}T\phi^{\star3}(x_2)\phi(x_1)\Big( \id -\frac {\ri \gl}{4!} \int d^4z \, \phi ^{\star 4} (z) \Big)\ket{0}_c
\,. \label{eq:verify-nc-qap}
\end{align}
In order to check explicitly that this relation actually holds to order $\gl$, all possible Wick contractions leading to connected graphs
have to be considered.
When working in $x$-space, particular care must be exercised concerning the order
of the resulting propagators due to the star products, for instance
\begin{align}
\contraction{\phi(x)\star}{\phi(x)}{\star\phi(x)\,}{\phi(y)}
\phi(x)\star\phi(x)\star\phi(x)\,\phi(y)&=\phi(x)\star\ri \prop(x-y)\star \phi(x)
\,.
\end{align}
Keeping this fact in mind, the explicit check of relation \eqnref{eq:verify-nc-qap} is straightforward, albeit tedious.

\section{BPHZ for {\nc} field theories}
\label{sec:BPHZ-nc}
%%%%%%%%%%%%%%%%%%%%%%%%%%%%%%%%%%%%%%%%%%

We recall~\cite{Das:2006, Piguet:1995} that the BPHZ method is a recursive subtraction scheme which does not
refer to or use any particular
regularization procedure: the renormalized Feynman graphs are obtained by subtracting out the divergent parts
of the integrands for all graphs describing divergent integrals.
These subtractions are equivalent to the addition of local, regularization dependent counterterms
to the Lagrangian, but there is no need to introduce these counterterms explicitly.

\subsection{General considerations}
\label{sec:bphz}
%==============================================================================

In {\nc} QFTs we encounter the problem that the usual BPHZ momentum space
subtractions do not eliminate the
infrared divergences which are due to {\uim}.
This may be illustrated by the
example of the {\nc} $\phi^4$-theory in Euclidean space whose Feynman rules are given by\footnote{The Minkowskian counterpart of these rules has already been given in \eqnref{eq:feynmanrules-naivemodel}.}
\begin{align}
&
\prop (p) = \inv{ p^2 + m^2 } \, ,
\nn \\
&
 V (p_1,p_2,p_3,p_4) \equiv - \l f( p_i, \th )
\nn \\
& = \frac{-\l}{3}
\left[
\cos \left( \tfrac{p_1\p_2}{2} \right)
\cos \left( \tfrac{p_3\p_4}{2} \right)
+\cos \left( \tfrac{p_1\p_3}{2} \right)
 \cos \left( \tfrac{p_2\p_4}{2} \right)
+\cos \left( \tfrac{p_1\p_4}{2} \right)
 \cos \left( \tfrac{p_2\p_3}{2} \right)
\right]
\,.
\label{eq:feynmanrules-naivemodel-euclidean}
\end{align}
For the non-planar part of the one-loop four-point $1PI$-graph, the BPHZ-subtraction in momentum space yields an integral of the following type:
\begin{align}
\label{bphzsub}
\hat J(p) \equiv
 \intk\left( \frac{\cos(k\p)}{((k+p)^2+m^2)(k^2+m^2)} - \frac{1}{(k^2+m^2)^2} \right)
 \,.
\end{align}
Here, $p$ denotes the total external momentum, $\tilde p_{\mu} \equiv \theta_{\mu \nu} p^{\nu}$,
and the second term in the integral (\ref{bphzsub})
represents the subtraction of the integrand evaluated at $p=0$.
(For later reference we note that $p\tilde{p} = p^{\mu} \theta_{\mu \nu} p ^{\nu} =0$.)
In commutative  $\phi^4$-theory, the phase factor $\cos(k\p)$  is absent and
the second term then cancels the logarithmic UV singularity of the first integral leaving just finite terms.
In {\nc} $\phi^4$-theory, the presence of the phase factor ensures that
the first term is UV finite, but it now suffers from an
infrared singularity in the external momentum which is \emph{not} canceled by the second term -- worse still, the second term introduces an additional UV singularity.

As a second example, consider the BPHZ-subtraction for the combined contributions of planar and non-planar parts:
\begin{align}
\hat J_{tot}(p) \equiv
 \intk\left( \frac{A+B\cos(k\p)}{((k+p)^2+m^2)(k^2+m^2)} - \frac{A+B}{(k^2+m^2)^2} \right)
 \,.
\end{align}
Clearly, the second term
(term proportional to $A+B$)
overcompensates the UV divergence of the
very first integral (term proportional to $A$)
while the infrared divergence remains present.

From these examples we conclude that the usual BPHZ-strategy of subtracting terms with $p=0$
from the integrand does not work for {\nc} theories.
Here, we suggest the following \emph{procedure for the BPHZ momentum space subtraction in {\nc} theories:}
consider $p$ and $\p$ as independent variables (though satisfying $p \tilde p =0$) when applying the subtraction rules of the BPHZ scheme,
i.e. subtract from the integrand terms with $p=0$
while maintaining the phase factors: instead of the subtraction (\ref{bphzsub}) we thus consider
\begin{align}
\label{checkJ}
\check J (p) \equiv
\intk\left( \frac{\cos(k\p)}{((k+p)^2+m^2)(k^2+m^2)} - \frac{\cos(k\p)}{(k^2+m^2)^2} \right)
 \,.
\end{align}
By proceeding in
this way, the logarithmic IR divergence of the first term is canceled by the second.

Expression \eqref{checkJ} is explicitly evaluated in \appref{app:bphz-example} leading to an integral over
a modified Bessel function $K_1(z)$ of second kind:
\begin{align}
\check J (p )  &=
\pi ^2 p^2 \int_0^1d\xi \ \frac{ (1-2 \xi )(\xi-1) \; K_1\left(\sqrt{\left(m^2-(\xi -1)
   \xi  p^2\right) \tilde{p}^2}\right)}{\sqrt{\frac{m^2- (\xi -1) \xi p^2}{\tilde{p}^2}}}
\,.
\end{align}
This expression is perfectly finite since the series expansion $K_1(z)=\inv{z} +\inv2 z \, {\rm ln} \, z + \mathcal{O}(z)$ implies that
we have for small values of $\p$:
\begin{align}\label{non_planar-result}
\check J (p )
% &= -2 \pi ^2 \left(\sqrt{1+\frac{4 m^2}{p^2}} \, \tanh
%    ^{-1}\left(\sqrt{\frac{p^2}{4 m^2+p^2}}\right)-1\right)+\cO(\p^2)\nn\\
&=- \pi ^2 \sqrt{1+\frac{4 m^2}{p^2}} \, \ln {\left[\frac{\sqrt{p^2+4 m^2}+\sqrt{p^2}}{\sqrt{p^2+4 m^2}-\sqrt{p^2}}\right]}+2\pi ^2+\cO(\p^2)
\,.
\end{align}
This expression is regular in the limit $p^2\to 0$.

Next, we consider the \emph{planar part} of the momentum space integral which is well known in the literature
since it also appears in the commutative $\phi^4$-theory~\cite{Piguet:1995}:
\begin{align}
\hat{J}_{pl}  (p)
= \intk\left( \frac{1}{((k+p)^2+m^2)(k^2+m^2)} - \frac{1}{(k^2+m^2)^2} \right)
 \,.
\end{align}
The application of the procedure described above does not require any extra
calculation:
By setting $\eta =0$ in \eqnref{Jpeta},
 we end up with
\begin{align}
\hat{J}_{pl}  (p)
=
 \pi^2p^2\int_0^1\! d\xi\, (1-2\xi)(\xi-1) \int_0^\infty\! d\l\,  \re^{-\l\left[m^2 -(\xi-1)\xi p^2\right]}
 \,.
\end{align}
After carrying out the integrations, we recover the well-known expression for the planar integral:
\begin{align}
\hat{J}_{pl}  (p)
=
- \pi ^2 \sqrt{1+\frac{4 m^2}{p^2}} \, \ln {\left[\frac{\sqrt{p^2+4 m^2}+\sqrt{p^2}}{\sqrt{p^2+4 m^2}-\sqrt{p^2}}\right]} +2\pi ^2
\,.
\end{align}
Comparing this result with expression~\eqref{non_planar-result} which holds for small values of $\p$, we conclude that our procedure for applying the BPHZ subtraction
leads to differences between the planar and non-planar parts which only appear in the higher-order terms $\cO(\p^2)$.

A general one-loop integrand $I(p_i,\p_i,k)$ with superficial degree of divergence $n$ (upon integration over $k$) should be replaced by
\begin{align}
R(p_i,\p_i,k)&=\left(1-t^n_p\right)I(p_i,\p_i,k)\,, \nn\\
(t^n_p f)(p_i,\p_i)
&:=f(0,\p_i)+\sum_jp^\m_j\left(\diff{}{p_j^\m}f(p_i,\p_i)\right)\Big|_{p_i=0}+ \ldots \nn\\
&\quad +\inv{n!}\sum_{j_1, \dots, j_n}p_{j_1}^{\m_1}\ldots p_{i_n}^{\m_n}\left(\diff{}{p_{j_1}^{\m_1}}\ldots \diff{}{p_{j_n}^{\m_n}}f(p_i,\p_i)\right)\Big|_{p_i=0}
\,.
\end{align}
Note that only the external momenta $p_i$
are involved in this subtraction scheme and that $\p_i\neq0$ at all stages so as to ensure convergence
of integrals as in the commutative case.
This also means, that one must exercise care in rewriting phase factors as $k\th p_i=k\p_i$ rather than $-\k p_i$.
Also note that phases not involving $k$ will not regularize the integrals and hence not lead to {\uim}.
There is, however, some ambiguity concerning our treatment of phases involving $p_i\th p_j$.
We hence suggest to rewrite these as $p_i\th p_j=\inv2(p_i\p_j-\p_ip_j)$.

The strategy outlined above, which amounts to a simultaneous subtraction of UV and IR divergences, allows us to overcome the  {\uim} problem at one-loop order.
This strategy seems fairly general and it should also be applicable to other theories, such as gauge theories.
In the next subsection, we will show for the example of non-commutative $\phi^4$-theory
that this procedure amounts to a one-loop order redefinition of the parameters appearing in the classical action.
In Subsection~\ref{higherloops}, we will comment on higher-order loops.

\subsection{One-loop renormalization of {\nc} \texorpdfstring{$\phi^4$}{phi**4}-theory}
\label{sec:bphz2}
%==============================================================================

We now apply the subtraction scheme presented in the previous subsection to the {\nc} $\phi^4$-theory.
Thus, we consider the following generalized action in Euclidean space:
\begin{align}
\label{BPHZaction}
S
[\phi]&=\intx\left[
\inv2 \, (1+A)
\pa^\m\phi\pa_\m\phi+ \inv2 \, (m^2+B)
\phi^2+ \inv{4!} \, (\l+C )
%(\pa,\th))
\phi^{\star4}\right]
\,.
\end{align}
Here, the parameter $A$ is dimensionless, $B$ has the same dimension as $m^2$ and $C$ the same dimension as $\l$.
The Fourier transforms of the parameters $A,B$ and $C$, which we denote by $a,b$ and $c$, potentially depend on the momentum and on $\theta$.
The dependence of $c$ on the momentum is motivated by the fact that the star product leads to a momentum dependent coupling.
As usual in the BPHZ-scheme, $m$ and $\l$ are already the one-loop renormalized mass and coupling, respectively.
We consider $a$, $b$ and $c$ to be of order $\hbar$ (i.e. one-loop expressions), and
for $\hbar \equiv 1$ the Feynman rules in momentum space are given by \eqnref{eq:feynmanrules-naivemodel-euclidean}.
In the following calculation it will be essential to keep $\p^2\neq\th^2 p^2$, since we treat $p$ and $\p$ as
independent variables in the BPHZ subtraction scheme as discussed in the previous subsection.

Including its one-loop correction,
the \emph{vertex} $\G^{(4)}(p_1,p_2,p_3,p_4) \equiv \G^{(4)}(p_i)$ is given by
(see \eqnref{eq:feynmanrules-naivemodel-euclidean} and references~\cite{Micu:2000,Aref'eva:1999sn})
\begin{align}
&\G^{(4)}(p_i)
=\left(\l+c(p_i,\th)\right)f(p_i,\th) \nn\\
&\quad +\frac{\l^2}{9}\!\int\!\!\frac{d^4k}{(2\pi)^4}\Bigg(\!\inv{(k\!+\!p_1\!+\!p_2)^2+m^2}\inv{k^2+m^2}\Big(1+\tinv2\sum_{i=1}^4 e^{\ri k\p_i}+e^{\ri k(\p_1+\p_2)}+\tinv{4}\!\sum_{i=3,4} e^{\ri k(\p_1+\p_i)}\Big) \nn\\
&\quad\qquad +p_2\leftrightarrow p_3+p_2\leftrightarrow p_4\Bigg)
\,.
\label{oneloop4point}
\end{align}
Thus, the vertex correction consists of integrals of the same type as those discussed
in the previous subsection.

We now add and subtract the same term to expression (\ref{oneloop4point}):
\begin{align}
\label{vert}
&\G^{(4)}(p_i)
=(\l+c(p_i,\th))f(p_i,\th)
\nn\\
&\;\;
+ \frac{\l^2}{9} \!\int\!\!\frac{d^4k}{(2\pi)^4}\!\left[\!\left(\inv{(k+p_1+p_2)^2+m^2}\inv{k^2+m^2}-\inv{(k^2+m^2)^2}\right)\!\!\Big(1+\tinv2\sum_{i=1}^4 e^{\ri k\p_i}+\ldots\!\Big)\!+\ldots\!\right]
\nn\\
&\;\; +\frac{\l^2}{9} \int\!\!\frac{d^4k}{(2\pi)^4}\left[\inv{(k^2+m^2)^2}\Big(1+\tinv2\sum_{i=1}^4 e^{\ri k\p_i}+\ldots\Big)+\ldots\right]
\,.
\end{align}
The second line of this expression is finite --- let us call it
$\widetilde{\Delta}' (p_i,\th ) $.
For \emph{small} external momenta $p_i$,
one has $f(p_i,\th) \approx 1$ and we get~\cite{Micu:2000}
\begin{align}
\G^{(4)}(p_i)
&\approx \l+c(p_i,\th)+\widetilde{\Delta}'(p_i,\th )+\frac{2\l^2}9\int\!\!\frac{d^4k}{(2\pi)^4}\inv{(k^2+m^2)^2} \nn\\
&\quad -\frac{\l^2}{6(4\pi)^2}\left[\sum_{i=1}^4\ln(m^2\p_i^2)+\sum_{i=2}^4\ln(m^2(\p_1+\p_i)^2)\right]
\,,
\end{align}
i.e. we have logarithmic IR divergences.

Following the lines of the BPHZ scheme, we now split $c$ according to
\begin{align}
\label{splitc}
c(p_i,\th)=c'+c''(p_i,\th)+c_\infty
\, ,
\end{align}
where $c'$ is an arbitrary finite constant and
\begin{align}
c''&\equiv\frac{\l^2}{6(4\pi)^2}\left[\sum_{i=1}^4\ln(m^2\p_i^2)+\sum_{i=2}^4\ln(m^2(\p_1+\p_i)^2)\right] \,, \nn\\
c_\infty&\equiv-\frac{2\l^2}9\int\!\!\frac{d^4k}{(2\pi)^4}\inv{(k^2+m^2)^2}
\,.
\end{align}
The function $\G^{(4)}$ resulting from the decomposition (\ref{splitc})
is denoted by $\G^{(4)}_r$: for small $\tilde p _i$ it thus reads
\begin{align}
\G^{(4)}_r(p_i)&=\l+c'+\widetilde{\Delta}'(p_i,\th)
\,,
\end{align}
where the constant $c'$ is to be fixed at a convenient symmetry point (s.p.) by the condition
\begin{align}
\G^{(4)}_r (p_i)\big|_{\textrm{s.p.}}&=\l\,, \qquad
{\rm i.e.}
\ \; c'=-\widetilde{\Delta}'\big|_{\textrm{s.p.}}
\,. \label{eq:fix-coupling-at-symmpoint}
\end{align}
This defines the physical coupling at the symmetry point and at one-loop order.

We now turn to the case of generic (i.e. not necessarily small) external momenta.
From (\ref{vert}) we then conclude that
\begin{align}
\G^{(4)}(p_i)&= (\l+c(p_i,\!\th))f(p_i,\!\th)+\widetilde{\Delta}'+\frac{\l^2}9\!\int\!\!\frac{d^4k}{(2\pi)^4}\inv{(k^2+m^2)^2}
\Big[\Big( 1\!+\tinv{2}\sum_i e^{\ri k\p_i}+\ldots\Big)
+ \dots \Big]
, \nn\\
c'&=-(\inv{f}\widetilde{\Delta}' ) \big|_{\textrm{s.p.}} \,,
\qquad\qquad
c''=-\frac{\l^2}{9f}\int\!\!\frac{d^4k}{(2\pi)^4}\inv{(k^2+m^2)^2}
\Big[\Big(\tinv{2}\sum_i e^{\ri k\p_i}+\ldots\Big)
+ \dots \Big]
\,, \nn\\
c_\infty&=-\frac{2\l^2}{9f}\int\!\!\frac{d^4k}{(2\pi)^4}\inv{(k^2+m^2)^2}
\,,
\end{align}
henceforth
\begin{align}
\G^{(4)}_r(p_i)&=(\l+c')f(p_i,\th)+\widetilde{\Delta}'(p_i,\th)\,, \nn\\
\G^{(4)}_r(p_i)\big|_{\textrm{s.p.}}&=\l
f(p_i,\th)\big|_{\textrm{s.p.}}
\,.
\end{align}
The symmetry point is defined in terms of the Mandelstam variables $s,t,u$ and some \emph{normalization mass} $\mu$:
\begin{align}
s&=(p_1+p_2)^2\,, && t=(p_1+p_3)^2\,, && u=(p_1+p_4)^2 \,,\nn\\
p_i^2&=\m^2
\ \ \forall i
\,, &&p_ip_j=-\frac{\m^2}{3}
\ \ \forall i,j
\,, & \Rightarrow \quad & s=\frac{4\m^2}{3}
\,. \label{eq:symm-point}
\end{align}

Next, we consider the \emph{two-point function} at the one-loop level, $ \G^{(2)}(p,-p) \equiv \G^{(2)}(p^2)$, which is given by
\begin{align}
\G^{(2)}(p^2)
&=(1+a)p^2+m^2+b(p,\th)-\frac{\l}{6}\int\!\!\frac{d^4k}{(2\pi)^4}\frac{2+\cos(k\p)}{k^2+m^2}
\,. \label{eq:real-scalar-2p}
\end{align}
Since the latter integral only depends on $\tilde p$ and not on $p$, we consider
$a=0$ and we split $b$ according to
\begin{align}
%a&=0\,, \qquad
b(p,\th) & =b'+b'' (p,\th) +b_\infty \,,
\nn\\
b''  (p,\th)&\equiv
\frac{\l}{6}\int\!\!\frac{d^4k}{(2\pi)^4}\frac{\cos(k\p)}{k^2+m^2}=\frac{\l}{24 \pi^2} \sqrt{\frac{m^2}{\p^2}} K_1\left(\sqrt{m^2 \p^2}\right)
\nn\\
&\approx \frac{\lambda }{24 \pi ^2 \p^2}
+\frac{m^2 \l}{96 \pi^2} \left[ \ln  \left(m^2\p^2/4\right)+2 \g_E-1\right]+\cO\left(\p^2\right) \,,
\label{expandsmall}
\nn\\
b_\infty&\equiv\frac{\l}{6}\int\!\!\frac{d^4k}{(2\pi)^4}\frac{2}{k^2+m^2}
\,,
\end{align}
where $b'$ is a finite constant.
The one-loop renormalized two-point function hence reads
\begin{align}
\G^{(2)}_r(p^2)&=p^2+m^2+b'
\,,
\end{align}
supplemented by the normalization conditions
\begin{align}
\label{normalprop}
\G^{(2)}_r(p^2)
\big|_{p^2=-m^2} &=0\,, \qquad
{\rm i.e.}\ \;
 b'=0 \,, \nn\\
\frac{d}{dp^2}\G_r ^{(2)}(p^2)\big|_{p^2=\m^2}=1
\,.
\end{align}

The calculation and reasoning presented in this subsection should also apply to  gauge theories
(in particular to the translation invariant models introduced in reference~\cite{Blaschke:2008a}), although the following complications are to be expected
in this case.

In Minkowski space, as long as time commutes with the spatial
coordinates (i.e. $\th^{i0}=0$), the on-shell condition $p^2=0$ does \emph{not} mean $\p^2=0$ since there is no time-component in $\p$. Hence, IR issues
\emph{only} arise in integrals,
but not in connection with the normalization conditions which are to be considered in
the actual renormalization process.

In Euclidean space, the on-shell condition $p^2=0$ may imply $\p^2=0$ depending on the structure and rank of $(\th^{\m\n})$.
Hence it
might be preferable to consider Min$(p^2+1/\p^2)$ as the renormalization point instead of $p^2=0$.
This would apply also to the normalization conditions fixing the physical parameters.

Before proceeding further, it is worthwhile to come back once more to the well-known non-planar contribution to the propagator
as described by the integral
\[b''  (p,\th)  \equiv
\frac{\l}{6} \int\!\!\frac{d^4k}{(2\pi)^4}\frac{\cos(k\p)}{k^2+m^2}\,,\]
see expression \eqref{expandsmall}.
For small values of $\tilde p ^2$, this integral exhibits a quadratic and a logarithmic IR divergence at $\tilde p ^2 =0$.
These IR divergences, which are tied to the UV divergences, represent the most basic manifestation
of the {\uim} problem appearing in non-commutative QFT. Multiple insertions of this non-planar diagram into a non-planar tadpole graph
lead to IR problems for higher order graphs~\cite{Minwalla:1999px, Micu:2000, Blaschke:2008b}.
These problems motivated the authors of reference~\cite{Rivasseau:2008a} to introduce the most singular (i.e. quadratic) divergence
into the classical Lagrangian by means of the non-local term
\begin{align}
\label{nonlocalterm}
\intx\left(\phi\frac{a'^2}{\wsq}\phi\right)
\,, \qquad \text{with }\; \wsq\equiv\widetilde{\pa}^\m\widetilde{\pa}_\m=\th^{\m\m'}\th_\m^{\ \n'}\pa_{\m'}\pa_{\n'}
\,,
\end{align}
where $a'$ represents a real dimensionless constant.
The renormalization of the resulting model was established using multi-scale analysis in reference~\cite{Rivasseau:2008a}
and the one-loop renormalization was discussed more specifically in reference~\cite{Blaschke:2008b}. The crucial point is that the
modified propagator $\prop \equiv (p^2 +m^2 + a'^2 /\p^2)^{-1}$
has a damping behavior for vanishing momentum (i.e. $\lim_{p\to 0} \prop (p) =0$)
which ultimately allows to cure the IR problems at any order.

In the BPHZ approach considered in this section, the inclusion of the non-local term (\ref{nonlocalterm}) into the classical
action modifies the quadratic part of the functional (\ref{BPHZaction}) according to
\[
\inv2 \, (1+A)
\pa^\m \phi\pa_\m\phi + \inv2 \, (m^2+B)
\phi^2
\quad \leadsto \quad
\inv2 \, (1+A)
\pa^\m \phi\pa_\m\phi +
\inv{2} \,\phi \Big( m^2  + \frac{a'^2}{\widetilde{\Box}} +B\Big)  \phi
\, ,
\]
the normalization condition (\ref{normalprop}) becoming
\[
\G_r ^{(2)}(p^2)
\big|_{p^2+ m^2+ \frac{a'^2}{\p^2} =0 } =0
\, .
\]
Hence, the one-loop
renormalized two-point function
 (\ref{eq:real-scalar-2p})
 is modified according to
\begin{align}
\G^{(2)}(p^2)
&=(1+a)p^2+m^2+b(p,\th)+\frac{a'^2}{\p^2}-\frac{\l}{6}\int\!\!\frac{d^4k}{(2\pi)^4}\frac{2+\cos(k\p)}{k^2+m^2+\frac{a'^2}{\k^2}}
\,, \label{eq:real-scalar-2p-nl}
\end{align}
where $b(p,\th)$ can now be interpreted as the renormalization of the new parameter $a'$ -- see
reference~\cite{Blaschke:2008b}.

\subsection{Beyond one loop}\label{higherloops}
%%%%%%%%%%%%%%%%%%%%%%%%%%%%%%%

Our momentum space subtraction scheme allowed us to discard all divergences at one-loop order.
Potential complications arise in higher-loop graphs:
\begin{itemize}
\item
For all divergent subgraphs of a given graph, we need an unambiguous distinction between internal and external momenta
since the latter are treated as passive variables in our subtraction scheme.
A point which is related to this issue is the need to
ensure that Zimmermann's forest formula still allows for a consistent treatment of all divergences.
\item
In order to establish renormalizability,  it has to be shown that
the ambiguities in the subtraction procedure result in a finite number of counterterms (redefinition of a finite number of parameters
determining the Lagrangian).
\end{itemize}
Further investigation of these problems and in particular the tackling of two-loop graphs
is currently under study~\cite{wip}.
Here, we only emphasize the following point. Since the ``na\"ive'' $\phi^4$-theory described at the one-loop level by the action
(\ref{BPHZaction}) is not renormalizable, it appears to be clear that the
presented subtraction of divergences leads at some stage to the inclusion of some extra term in the classical Lagrangian.
In the next section, we will discuss the nature of non-local terms
which can appear in the framework of non-commutative QFT.
These arguments lead to the conclusion that the only non-local term which can be included in a translation invariant model
for a real scalar field with quartic self-interaction is the one given by expression~(\ref{nonlocalterm}).

%==============================================================================
\section{Nature of non-localities in {\nc} models}
%==============================================================================
\label{sec:locality}
The approach of algebraic renormalization has been developed for local field theories.
One would expect that
due to the loss of locality we have an infinite number of possible counterterms for any such theory.
However, this is not really the case since the non-locality only originates from the star product and only leads to IR divergent non-local counterterms in the effective action\footnote{We should mention at this point that already Filk~\cite{Filk:1996} realized that the UV structure of a renormalizable field theory is not affected by the non-locality induced by non-commutativity.}.
This is a consequence of the {\uim} and thereby these problematic counterterms can only have an IR divergence of degree equal or less than the superficial (maximal) degree of UV divergence $d_{\gamma}$.
Furthermore, these terms are generated by phases depending on the combination $\p^\m=\th^{\m\n}p_\n$ where $p$ is some external momentum.
Hence, IR divergent non-local terms $N(p)$ behave like
\begin{align}
\lim_{p\to0}
N(p) \sim\inv{(\p^2)^s}
\,,\qquad \text{with }
%s>0
0< s < \frac{d_{\gamma}}{2}
\,. \label{eq:IR-degree}
\end{align}

%%%%%%%%%%%%%%%%%%%%%%%%%%%%%%%%%%%%%%%%%%%%%%%%%%%%%%%%%%%%%%%%%%%%%
\subsection{Translation invariant models}

Let us consider $\phi^4$-theory in commutative space as an example for a power counting renormalizable theory.
Its superficial degree of UV divergence is given by $d_\g=4-E$ where $E$ denotes the number of external legs.
If we generalize this model to {\nc} space by replacing point-wise products of fields by star products, non-local counterterms arise.
But the degree of IR divergence, as introduced in \eqnref{eq:IR-degree}, can be at most $s=2-E/2$ due to power counting.
Moreover, if we use Schwinger's exponential parametrization for the integrals (see e.g. Eqns. (\ref{Jp}),(\ref{Jpeta})), we can use the phase factors
involving $\p$ to complete the squares in Gaussian-type integrals: for small values of $\p$, the resulting expressions behave as $1/(\p^2)^s$ where $s$ is
a multiple of $D/4$ ($D$ being the space-time dimension).
Thus, a combination $(\p^2)^{(2-E/2)}\times$(counterterm) should be \emph{local}.

This argument motivates us to generalize the algebraic renormalization procedure by allowing non-local counterterms which become local upon multiplication by $(\p^2)^s$ with $s=2 - \frac{E}{2}$ while excluding non-local terms of a different nature.
\begin{example}
Consider the real scalar field theory action in Euclidean space
\begin{align}
S=\intx\left(
\inv2 \,
(\pa_\m\phi)( \pa^\m\phi ) + \inv2 \, m^2\phi^2+\frac{\l}{4!}
\phi^{\star4}\right)
\,.
\end{align}
It is translation invariant as well as invariant under the transformation $\phi\to-\phi$.
In commutative space, i.e. for the ordinary point-wise product, the action
already includes all terms which are allowed for a local power-counting renormalizable theory
in four dimensions.
However, in {\nc} space one more additional term is allowed according to our considerations above, namely~(\ref{nonlocalterm}).
This term is invariant under translations as well as under the transformation $\phi\to-\phi$.
Furthermore, it has mass dimension 4 (which is the maximal value allowed for the 4-dimensional model under consideration), and it has an IR divergence of order $1/\p^2$ which is the maximum degree allowed by power counting since $s=1$ for a counterterm with two fields $\phi$.
\end{example}

\begin{example}
Now consider the complex scalar field action in Euclidean space\footnote{Note that only the vertex proportional to $\l_2$ exhibits {\uim} -- at least at the one-loop level -- as was shown in references~\cite{Aref'eva:2000hq}.}
%,Arefeva:2000uu,Aref'eva:2000bg}.}
\begin{align}
\G^{(0)}=\intx\left(\pa_\m\bar\varphi\pa^\m\varphi+\bar\varphi\frac{a'^2}{\wsq}\varphi+m^2\bar\varphi\varphi+\frac{\l_1}{4}\, \bar\varphi\star\varphi\star\bar\varphi\star\varphi \, + \frac{\l_2}{4}\, \bar\varphi\star\bar\varphi\star\varphi\star\varphi \right)
\,. \label{eq:action-bsp2}
\end{align}
This action is again translation invariant, invariant under $\varphi\to-\varphi$, $\bar\varphi\to-\bar\varphi$, invariant under the charge conjugation transformation $\mathcal{C}\,:\ \varphi\leftrightarrow\bar\varphi$.
Furthermore, there exists a continuous internal symmetry transformation leaving the action \eqref{eq:action-bsp2} invariant, namely the
global $U(1)$ transformation
\begin{align}
\d_\vare\varphi&=\ri\vare\varphi\,, \qquad\qquad
\d_\vare\bar\varphi=-\ri\vare\bar\varphi
\,,
\end{align}
the latter leading to the classical Ward identity
\begin{align}
\mathcal{W}\G^{(0)}&=0 \,, &
\text{with }\;
\mathcal{W}&=\intx\left(\varphi\var{\ }{\varphi}-\bar\varphi\var{\ }{\bar\varphi}\right)
\,.
\end{align}
Note that the only possible non-local term which is allowed by power counting and compatible with the symmetries,
namely $\bar\varphi\frac{a'^2}{\wsq}\varphi$ has already been included in the action \eqref{eq:action-bsp2}.

Assuming the quantum action principle to hold in the {\nc} case and including non-local terms of the nature described above, we
can follow the standard procedure for studying the breaking of symmetry by radiative corrections (see reference~\cite{Piguet:1995}
and Eqns. (\ref{W2})-(\ref{Wariations}) below).
We then find the following basis of integrated insertions of dimension
less or equal to four:
\begin{align}
\mathcal{W}\G^{(n-1)}&=\hbar^n \IDelta
+ \mathcal{O} (\hbar ^{n+1})
\,, \nn\\
\IDelta :\quad &\intx\left(\varphi^2-\bar\varphi^2\right)\,, &&\intx\left(\pa_\m\varphi\pa^\m\varphi-\pa_\m\bar\varphi\pa^\m\bar\varphi\right)\,, &&\intx\left(\varphi\inv{\wsq}\varphi-\bar\varphi\inv{\wsq}\bar\varphi\right)\,, \nn\\
&\intx\left(\varphi^{\star4}-\bar\varphi^{\star4}\right)\,, &&\intx\left(\varphi^{\star3}\bar\varphi-\bar\varphi^{\star3}\varphi\right)
\,. \label{eq:qap-basis-complex}
\end{align}
Thus, the basis of admissible terms includes one
non-local  expression.
\end{example}

\paragraph{Remark.}
%%%%%%%%%%%%%%%%%%%%
The additional non-local terms which are bilinear in the fields will lead to damping properties of the field propagators in the IR regime.
These terms should take care of the IR problems  in higher loop integrals (see next section).
However, it must be shown explicitly that they are sufficient to render the model under consideration renormalizable.

\subsection{Non-translation invariant models}

As remarked in Subsection~\ref{briefreview},  non-local terms of the type ``product of traces'',
e.g.  $\inv{\th ^2} ({\rm Tr} \, \phi )^2$ or $ ({\rm Tr} \, \phi ^2)^2$,
appear to be natural on {\nc} spaces~\cite{Grosse:2008df}.
In particular, a term of this type ensures the renormalizability of the harmonic model for the
$\phi^4$-theory on degenerate Euclidean Moyal space -- see Eqn. (\ref{grossevignes}).
Yet, it would be preferable to have a more precise characterization of the generic non-local terms
which are admissible on Moyal space.

\section{Algebraic renormalization of NCQFTs}
\label{sec:AR-nc}
%%%%%%%%%%%%%%%%%%%%%%%%%%%%%%%%%%

The quantization of Lagrangian models with continuous (rigid or local) symmetries can conveniently be addressed using the approach
of algebraic renormalization~\cite{Piguet:1995}. In particular, the BPHZ subtraction scheme can be applied
to these models.

\subsection{The self-interacting complex scalar field with a rigid \texorpdfstring{$U(1)$}{U(1)} symmetry}
\label{sec:complex}
%==============================================================================

In this section we consider models in four dimensional
 {\nc} Euclidean space for a complex scalar field $\vp$ with a self-interaction
and we try to apply the methods of algebraic renormalization.
More precisely, we consider models with a rigid  $U(1)$ symmetry,
i.e. invariance under the infinitesimal phase transformations
\begin{align}
\label{rigidtransf}
\delta _{\varepsilon} \vp
=  \ri \varepsilon \vp  \, ,
\qquad
\delta _{\varepsilon} \vpb
  = - \ri \varepsilon \vpb  \, ,
\end{align}
where $\varepsilon$ denotes a constant infinitesimal real parameter.

As noted some time ago by Aref'eva et al.~\cite{Aref'eva:2000hq},
%,Arefeva:2000uu,Aref'eva:2000bg},
the star product allows for two different
$U(1)$-invariant quartic interactions:
\begin{align}
\label{naive}
S_0[\varphi, \bar{\varphi} ] \equiv
\intx\left[ (\pa_\m \bar\varphi ) ( \pa^\m\varphi)  +m^2 \bar\varphi\varphi
+ \frac{\l_1}{4}  \, \bar\varphi \star\varphi \, \vpb \star \varphi
+ \frac{\l_2}{4} \, \bar\varphi \star \bar\varphi \, \varphi \star \varphi \right]
\,.
\end{align}
Here, $\lambda_1$ and $\lambda_2$ are two independent coupling constants and
we have used the properties of the star product to drop one star in the quartic terms.
The action \eqref{naive}, which will be referred to as the ``na\"ive'' model,
is invariant under translations, under the rigid $U(1)$ transformations (\ref{rigidtransf}),
as well as under the discrete transformations of charge conjugation $\mathcal{C}\,:\ \varphi\leftrightarrow\bar\varphi$
and of reflection $\varphi\to-\varphi, \, \bar\varphi\to-\bar\varphi$.
The corresponding Feynman rules have been given in references~\cite{Aref'eva:2000hq}.
%,Arefeva:2000uu,Aref'eva:2000bg}.
Since the bilinear terms of the action do not involve the star product, the propagator
in momentum space is the same as in the commutative theory, i.e. $(p^2 +m^2)^{-1}$.
In momentum space the interaction terms read\footnote{Note that our convention for the star product differs by a factor $1/2$ from the one used in ref.~\cite{Aref'eva:2000hq},
%,Arefeva:2000uu,Aref'eva:2000bg},
hence the factors $1/2$ in the cosines below.}
 \begin{align}
 \label{momint}
 \int d^4x \, V(\vp, \vpb)
 &= \frac{-1}{4(2 \pi )^4} \int d^4p_1 \cdots d^4p_4
 \, \delta ^{(4)} \left( \sum_{i=1}^4 p_i \right) \,
 \tilde{\vpb} (p_1)  \tilde{\vp} (p_2)  \tilde{\vpb} (p_3)  \tilde{\vp} (p_4)
% \left\{ \l_1 \left[
% \cos \left( {p_1\p_2} \right) \cos \left( {p_3\p_4} \right)
 %\right. \right.
  \nn \\
&\quad \times \left[ \l_1 \cos \left( \frac{p_1\p_2}2 + \frac{p_3\p_4}2 \right)
 +
 \l_2 \cos\left( \frac{p_1\p_3}2 \right)\cos\left( \frac{p_2\p_4}2 \right)
\right]
%& \   \left.
%\left. + \cos \left( {p_1\p_3}\right) \cos \left( {p_2\p_4} \right)  \right]
 % +
 %\l_2 \left[ \cos\left( {p_1\p_2} \right)\cos\left( {p_3\p_4} \right) \right]
 %\right\}
 %\tilde{\vpb} (p_1)  \tilde{\vp} (p_2)  \tilde{\vpb} (p_3)  \tilde{\vp} (p_4).
\,.
\end{align}

%%%%%%%%%%%%%%%%%%%%%%%%%%%%%%%%%%%%%%%%%%%%%%%%%%%%%%%%%%%%%%%%%%%%%%%%%%%%%%%%%%%%%%%%%%%%%%%%%%%%%%%%%%%%%%
\subsection{Self-interaction \texorpdfstring{$\bar\varphi \star\varphi \star \vpb \star \varphi$}{barphi-phi-barphi-phi}}

To start with, we consider the na\"ive model (\ref{naive}) with $\l_2 =0$, i.e. for
a quartic term with ``alternating" fields:
 \begin{align}
\label{naivealtern}
S_{al} [\varphi, \bar{\varphi} ]
\equiv \intx\left[ (\pa_\m \bar\varphi ) ( \pa^\m\varphi)  +m^2 \bar\varphi\varphi
+
\frac{\l_1}{4}
\, \bar\varphi \star \varphi \star \vpb \star \varphi
\right]
\,.
\end{align}
It was shown in references~\cite{Aref'eva:2000hq}
%,Arefeva:2000uu,Aref'eva:2000bg}
that this model is renormalizable at the one-loop level.
Indeed the considered interaction does not lead to a {\uim} -- at least at the one-loop level -- since the interaction (\ref{momint}) with $\lambda_2=0$
does not yield an infrared problematic tadpole contribution at the one-loop level.
Later it was argued~\cite{Chepelev:2000} that the model is renormalizable to all orders.

The following extension of the action (\ref{naivealtern}) has also been studied in the literature:
\begin{align}
\label{GWLSZ}
S_{GWLSZ}
[\varphi, \bar{\varphi} ]
\equiv & \intx \left[ ( \overline{D_{\mu} \varphi }) \star (D^{\mu} \varphi ) +m^2 \bar\varphi\varphi
+ \Omega ^2 \, (\overline{\tilde{x}_{\mu} \varphi }) \star ( \tilde{x}^{\mu} \varphi )
+ \frac{\l_1}{4} \, \bar\varphi \star \varphi \star \vpb \star \varphi \right] .
\end{align}
Here, $D_{\mu} \varphi \equiv \pa_{\mu}  \varphi - \ri \alpha  \tilde{x}_{\mu} \star \varphi$
can be viewed as the covariant derivative describing
the coupling to a constant ``magnetic field" and the term in $\Omega^2$ is the complex version of the
Grosse-Wulkenhaar harmonic term.
The model (\ref{GWLSZ}) has been proven to be renormalizable to all orders~\cite{Gurau:2005gd}
and the model with $\Omega =0$ and $\alpha \neq 0$ is known as the {\it Langmann-Szabo-Zarembo (LSZ) model,} this model being exactly solvable
for $\alpha =1$ \cite{Langmann:2003if}.

%%%%%%%%%%%%%%%%%%%%%%%%%%%%%%%%%%%%%%%%%%%%%%%%%%%%%%%%%%%%%%%%%%%%%%%%%%%%%%%%%%%%%%%%%%%%%%%%%%%%%%%%%%%%%%
\subsection{Extended action for the complex scalar field}

% \paragraph{The model.}

As discussed in references~\cite{Aref'eva:2000hq},
%,Arefeva:2000uu,Aref'eva:2000bg},
the na\"ive model (\ref{naive}) is not renormalizable for arbitrary values of $\l_1$
and $\l_2$, although it is in the particular cases $\l_2 =0$ and $\l_1 = \l_2$.
In order to have a renormalizable interaction for a generic value\footnote{As specified in the introduction, we only consider non-supersymmetric models, but it should be noted that supersymmetry would also render the model with $\l_2 \neq 0$ renormalizable~\cite{Girotti:2000}.} $\l_2 \neq 0$, we include an extra non-local term into the action of the type
considered earlier\footnote{In order to simplify computations, we consider the matrix $(\th^{\m\n})$ to be block-diagonal so that $\k^2=\th^2k^2$,
see Eqns. \eqref{Gurauaction}, \eqref{blockdiag}.}
for a real scalar field~\cite{Rivasseau:2008a, Blaschke:2008a, Blaschke:2008b}:
\begin{align}
\G^{(0)} [\vp, \vpb ]
\equiv
\intx \left[ (\pa_\m\bar\varphi) ( \pa^\m\varphi ) + \bar\varphi\, \frac{a^2}{\square} \, \varphi +m^2\bar\varphi\varphi
+ \frac{\l_1}{4}  \, \bar\varphi \star\varphi \star \vpb \star \varphi
+\frac{\l_2}{4} \, \bar\varphi\star\bar\varphi\star\varphi\star\varphi \right]
\,.
\label{ext-action}
\end{align}
Here,  the parameter $a$ is assumed to have the form $a=a'/\th$ where $a'$ represents a real dimensionless constant.
The propagator in momentum space reads
\begin{align}
%\raisebox{-1.5pt}[1pt][0pt]{\includegraphics[scale=0.8, trim=0 0 0 10,clip=true]{prop.eps}}= \; G(k)=
\prop (k) = \inv{k^2+m^2+\frac{a^2}{k^2}}\,,
\label{eq:propagator}
\end{align}
and its ``damping'' behavior~\cite{Rivasseau:2008a} for
vanishing momentum (i.e. $\lim\limits_{k \to 0} \prop (k)  =0$)
allows us to avoid potential IR divergences in higher loop graphs~\cite{Blaschke:2008b}.
The vertex in momentum space can be read off from \eqnref{momint}.

The \emph{one-loop corrections to the propagator} (including a symmetry factor of $1/2$) are given by
\begin{align}
\Pi(p)=  \frac{-1}{4} \int\limits_{\mathbb{R} ^4} \frac{d^4k}{(2\pi)^4} \,
\frac{(\l_1+\l_2) +\l_2\cos(k\p)}{k^2+m^2+\frac{a^2}{k^2}} \equiv
\Pi^{\text{plan}}+\Pi^{\text{n-pl}}(p)\, ,
\label{eq:loopint-1}
\end{align}
where $\Pi^{\text{plan}}$ and $\Pi^{\text{n-pl}}$ denote the \emph{planar} and \emph{non-planar parts}, respectively.
In comparison to the case of a real scalar field, the numerical prefactors of planar and non-planar parts have changed.

The integral (\ref{eq:loopint-1}) has been evaluated  in reference~\cite{Blaschke:2008b} by using
Schwinger's exponential parametrization, the decomposition $\cos(k\p)= \inv{2}\sum\limits_{\eta=\pm1}
{\rm e} ^{\ri\eta k\p}$ and
\begin{align}
\inv{k^2+m^2+\frac{a^2}{k^2}}  = \
&\frac{k^2}{\left(k^2+\frac{m^2}{2}\right)^2-M^4}=
\inv{2}\sum\limits_{\zeta=\pm1}\frac{1+\zeta\frac{m^2}{2M^2}}{k^2+\frac{m^2}{2}+\zeta M^2}\, ,
\label{eq:facdec}
\end{align}
where $M^2\equiv\sqrt{\frac{m^4}{4}-a^2}$ (which may be real or purely imaginary depending on the value of the parameter $a$).
For $\p ^{\, 2} \ll 1$, the non-planar part behaves like
\begin{align}
\label{smallp}
\Pi^{\text{n-pl}}(p)
&=\frac{-\l_2}{4(4\pi)^2} \Bigg[ \frac{4}{\p^{\, 2}} + m^2 \ln
\left(\p^{\, 2} \sqrt{\tfrac{m^4}{4}-M^4}\right) %\nonumber \\
% &\quad\hspace{2.3cm}
+\left(M^2+\tfrac{m^4}{4M^2}\right)\ln\sqrt{\frac{\frac{m^2}{2}+M^2}{\frac{m^2}{2}-M^2}}\, \Bigg] +\mathcal{O}(1) \, ,
\end{align}
and thereby involves a quadratic IR divergence (and a subleading logarithmic IR divergence). For $a\to0$ (i.e. $M^2\to\frac{m^2}{2}$) this result reduces to the one for the na\"ive model, i.e.
\begin{align}
 \lim_{a\to0}\Pi^{\text{n-pl}}(p)
&=\frac{-\l_2}{4(4\pi)^2} \left[ \frac{4}{\p^{\, 2}} + m^2 \ln\left(\p^2 m^2\right) \right] +\mathcal{O}(1)
\,.
\end{align}

The integral defining the \emph{planar part} does not contain a phase factor  and is therefore UV divergent. It can be regularized by introducing a cutoff $\Lambda$ and subsequently taking the limit $\p^{\, 2} \to0$, as explained in reference~\cite{Blaschke:2008b}.
The final result can be expanded for large values of $\Lambda$, yielding
\begin{align}\label{eq:PI_planar_result}
\left( \Pi^{\text{plan}} \right)_{{\rm regul.}} (\Lambda )
&= \frac{-(\l_1+\l_2)}{4(4\pi)^2} \Bigg[ 4\Lambda^2+m^2\ln
\left(\tinv{\Lambda^2}\sqrt{\tfrac{m^4}{4}-M^4}\, \right)
\nonumber\\
&\quad\hspace{2.3cm} +\left(M^2+\tfrac{m^4}{4M^2}\right)
\ln\sqrt{\frac{\frac{m^2}{2}+M^2}{\frac{m^2}{2}-M^2}} \, \Bigg]
+\mathcal{O}(1).
\end{align}

The one-loop renormalization can be discussed
following the case of a real scalar field, see reference~\cite{Blaschke:2008b} or \secref{sec:bphz2}.
Accordingly, the action (\ref{ext-action}) is expected to be stable with respect to radiative corrections
to all orders.
%and an explicit may proceed for instance along the lines of the multi-scale analysis~\cite{Rivasseau:2008a}.

We now turn to the symmetries of the theory.
The invariance of the tree level action (\ref{ext-action})
under rigid $U(1)$-transformations is expressed by the classical
Ward identity
\begin{align}
\label{W2}
\mathcal{W}\Gamma^{(0)}[\varphi,\bar\varphi]=0 \, ,
\qquad {\rm with} \ \
\mathcal{W}& \equiv \intx\left(\varphi\var{\ }{\varphi}-\bar\varphi\var{\ }{\bar\varphi}\right)
\,.
\end{align}
%In order to discuss the symmetry content of the model, we rewrite the one-loop correction to the propagator as given by
%(\ref{eq:loopint-1}) in the integrated form
%\begin{align}
%\Delta =  \int \frac{d^4p}{(2\pi)^4} \,
%\tilde{\vp} (p) \,
%\left[ \Pi^{\text{plan}}+\Pi^{\text{n-pl}}(p) \right] \, \tilde{\vpb} (-p)
%\, .
%\label{eq:loopint-int}
%\end{align}
%The invariance of this expression
%\begin{align}
%\mathcal{W} \Delta \equiv
%\int \frac{d^4p}{(2\pi)^4} \, \left( \tilde{\vp}  \var{\ }{\tilde{\vp} } - \tilde{\vpb} \var{\ }{\tilde{\vpb}} \right) \Delta =0
%\, .
%\label{eq:ward-id}
%\end{align}
The tools of algebraic renormalization allow us to investigate whether this symmetry holds to all orders of perturbation theory.
For this investigation we use the following notation for the loop expansion of the vertex functional:
\begin{align}
\Gamma_{(n-1)}=\sum_{m=0}^{n-1} \hbar^m\Gamma^{(m)}
\,.
\end{align}
%\begin{align}\label{eq:W1}
%\mathcal{W}&=\intx\left(\varphi\var{\ }{\varphi}-\bar\varphi\var{\ }{\bar\varphi}\right)
%\end{align}
Using the Ward operator $\mathcal{W}$
and the charge conjugation
$\mathcal{C}:\varphi\leftrightarrow\bar\varphi$, we now try to show recursively that the radiative corrections
will not destroy the rigid $U(1)$ symmetry of the vertex functional
at higher order.
At order $(n-1)$ in $\hbar$, we have
\begin{align}
\label{anomaly}
\mathcal{W}\Gamma_{(n-1)}=\hbar^n\Delta\cdot\Gamma=\hbar^n\Delta+\cO(\hbar^{n+1})
\,,
\end{align}
where we assumed that we can ensure  the symmetry with appropriate counterterms at the $(n-1)$ loop order.
In the last equality of Eqn. (\ref{anomaly}) we have assumed the validity of the QAP: the symmetry breaking term
 $\Delta$  then represents a Poincar\'e invariant, integrated insertion with a mass-dimension less or equal four.
In contrast to the commutative case (where $\Delta$ must be local), we presently
allow for
non-local insertions of the type discussed in \secref{sec:locality}.
We note that the expression $\Delta$ is odd with respect to charge conjugation
(i.e. $\mathcal{C} \Delta=-\Delta$)
as a consequence of the fact that the Ward operator $\mathcal{W}$
is also odd under the operation $\mathcal{C}$.

The classical field functionals $\Delta$ obeying all of the constraints we just listed can be expanded
with respect to the  basis given in
\eqnref{eq:qap-basis-complex}.
Obviously, the five basis elements can be written as the $\mathcal{W}$-variations of the following five
independent field functionals $\hat\Delta$:
\begin{align}
 &\inv2\intx\left(\varphi^2+\bar\varphi^2\right)\,, &\inv2\intx\left(\pa_\m\varphi\pa^\m\varphi+\pa_\m\bar\varphi\pa^\m\bar\varphi\right)\,, &&\inv2\intx\left(\varphi\inv{\wsq}\varphi+\bar\varphi\inv{\wsq}\bar\varphi\right)\,, \nn\\
&\inv4\intx\left(\varphi^{\star4}+\bar\varphi^{\star4}\right)\,, &\inv2\intx\left(\varphi^{\star3}\bar\varphi+\bar\varphi^{\star3}\varphi\right)
\label{Wariations}
\,.
\end{align}
Thus, one has
\begin{align}
\mathcal{W}\hat\Delta_i=\Delta_i
\,,
\end{align}
which implies  that we can rewrite \eqref{anomaly} as
\begin{align}
\mathcal{W}\Gamma_{(n-1)}=\hbar^n\mathcal{W}\hat\Delta+\cO(\hbar^{n+1})
\,, \qquad {\rm with} \ \  \hat\Delta= \sum_{i=1}^5  r_i\Delta_i
\, .
\end{align}
Let $S_{(n-1)}$ denote the action which includes all counterterms up to order $(n-1)$
and which leads to the vertex functional $\Gamma_{(n-1)}$.
If we now replace the action $S_{(n-1)}$ by the new action
\begin{align}
S_{(n)}=S_{(n-1)}-\hbar^n\hat\Delta
\,,
\end{align}
then we get the new vertex functional
\begin{align}
\Gamma_{(n)}=\Gamma_{(n-1)}-\hbar^n\hat\Delta
+ \mathcal{O} (\hbar^{n+1} )
\,,
\end{align}
so that \eqref{anomaly} yields
\begin{align}
\mathcal{W}\Gamma_{(n)}=0+\cO(\hbar^{n+1})
\,.
\end{align}
Thus, we have verified recursively that the Ward identity holds to all orders
and that there is no anomalous breaking of symmetry.

We stress that the previous procedure can only be expected to work for models for which one has taken care of the {\uim} problem.
Thus, for the considered example
it is crucial to include the $1/\p^2$-term in the action which implements the IR damping in the propagator.

%==============================================================================
\section{Conclusions and final remarks}
%==============================================================================
\label{sec:conclusions}

In this paper we have shown the validity of the QAP
at the lowest orders of perturbation theory for the case of a
scalar field theory on {\nc} Minkowski space.
We did not encounter potential obstructions or inconsistencies brought about the non-commutativity.
Accordingly, we expect the QAP to hold in general for renormalizable models, but a complete proof still remains to be elaborated.

In the sequel, we have discussed the difficulties that one encounters when applying the usual BPHZ approach to
field theories on {\nc} Euclidean space.
In the commutative case, the usual BPHZ subtraction scheme together with Zimmermann's forest
formula~\cite{Bogoliubov:1957gp}
%,Hepp:1966eg,Zimmermann:1969}
iteratively solves all problems raised by multi-loop graphs with overlapping divergences.
For the {\nc} case, we have put forward a procedure to remedy these problems and we have explicitly shown
that our strategy works at the one-loop level.
This result, which was not obvious a priori, represents a crucial first step
towards the definition of a complete and consistent scheme.
At higher-loop order, overlapping divergences represent a delicate issue for the renormalization
and in {\nc} field theories the  UV/IR mixing is at the origin of potential IR divergences.
As discussed in \secref{sec:BPHZ-nc},
the UV/IR mixing problem is overcome by the inclusion of the $1/p^2$-term into the action.
We will report elsewhere~\cite{wip} on
higher-loop results obtained by application of our substraction scheme
and in particular on the consistent treatment of
so-called sunset graph which is a typical two-loop graph with an overlapping divergence.

Finally, we have analyzed the nature of non-localities which can appear in typical {\nc} field theories and we applied the method of algebraic renormalization while taking into account these non-localities.
We hope that this program can be further completed so as to be applied to {\nc} gauge field theories
whose renormalizability represents an outstanding problem.

\section*{Acknowledgments}
%%%%%%%%%%%%%%%%%%%%%%%%%%%%%%%%%%%%%%%%%%%%%%%
The authors would like to express their gratitude to H. Grosse, O. Piguet and F. Vignes-Tourneret for valuable discussions.

D.N. Blaschke is a recipient of an APART fellowship of the Austrian Academy of Sciences.

We wish to thank the anonymous referees for their pertinent and
constructive comments which contributed to the clarification of several points,
as well as for pointing out several relevant references.

\appendix
\section{Evaluation of expression \texorpdfstring{\eqref{checkJ}}{(49)}}
\label{app:bphz-example}
%%%%%%%%%%%%%%%%%%%%%%
To evaluate expression \eqref{checkJ}, we write
\begin{align}
\cos(k\p)=\, & \inv{2}\sum\limits_{\eta=\pm1} \re ^{\ri\eta k\p}
 \, ,
\end{align}
and we use Schwinger's parametrization:
\begin{align}
\check J (p)  & =  -    \intk \; \frac{2kp+p^2}{((k+p)^2+m^2)(k^2+m^2)^2}
\; \cos(k\p)
=
 - \inv{2}\sum\limits_{\eta=\pm 1}  \, \check J_{\eta} (p ) \, ,
\label{Jp}
\end{align}
with
\begin{align}
\check J_{\eta} (p )  & =
\intk \left(2kp+p^2\right)\int\limits_0^\infty\! d\a\, \a\int\limits_0^\infty\! d\b \; \exp\! \left[ {-\a\left(k^2+m^2\right)-\b\left[(k+p)^2+m^2\right] + \ri\eta k\p } \right]
 . \label{Jpeta}
\end{align}
In order to carry out the $k$-integral in $\check J_{\eta} (p )$, we complete the square in the exponent:
\begin{align}
\check J_{\eta} (p )
 &= \int\limits_0^\infty\! d\a\,\a \int\limits_0^\infty\! d\b\int\! d^4k^\prime\left(2kp+p^2\right) \exp\! \left[ {-(\a+\b){k^\prime}^2 +\tfrac{(\b p - \frac{\ri}{2} \, \eta \p)^2}{\a+\b}-\b p^2-(\a+\b)m^2}
\right]
\end{align}
where
\begin{align}
 k^\prime=k+\frac{\b p - \frac{\ri}{2} \, \eta \p}{\a+\b}
 \,.
\end{align}
After expressing the factor $(2kp+p^2)$ in terms of $k ^{\prime}$, using $p \tilde p =0$,
and performing the Gaussian integration over $k ^{\prime}$, we get
\begin{align}
\check J_{\eta} (p )
%  &= -p^2 \int\limits_0^\infty\! d\a\,\a \int\limits_0^\infty\! d\b \left(\tfrac{2\b }{\a+\b}- 1 \right)  \exp\! \left[ \tfrac{(\b p - \frac{\ri}{2} \, \eta \p )^2}{\a+\b}-\b p^2-(\a+\b)m^2 \right]
%   \int\! d^4k^\prime \,  \re^{-(\a+\b){k^\prime}^2}
%  \nn\\
 &=- \pi^2p^2 \int\limits_0^\infty\! d\a \int\limits_0^\infty\! d\b \left(\frac{2\b}{\a+\b}-1\right)
 \frac{\a}{(\a+\b)^2}
\, \exp\! \left[ {\frac{(\b p - \frac{\ri}{2} \, \eta \p )^2}{\a+\b} -\b p^2-(\a+\b)m^2} \right].
\end{align}
The change of variables $(\a , \b ) \to (\lambda, \xi)$ defined by
\begin{align}
&  \a=(1-\xi)\l\,,\quad \b=\xi\l\,,\quad  {\rm with} \ \;  \lambda \in [  0 , \infty [ \, , \quad \xi \in [0,1 ]  \,,
\\
{\rm hence} \ \; & \a+\b=\l\,,\quad d\a d\b=\l d\l d\xi
\, ,
\nn
\end{align}
leads to
\begin{align}
\label{non_planar-integral}
\check J_{\eta} (p )
% & = -\pi^2 p^2 \int_0^1 \! d\xi \int_0^\infty \! d\l \, (2\xi-1) (1-\xi) \,
% \exp\! \left[ \inv{\lambda} \, (\xi \lambda p - \frac{\ri}{2} \, \eta \p )^2 - \xi \l p^2 - \l m^2  \right] \nn\\
 &=- \pi^2p^2\int\limits_0^1\! d\xi\, (1-2\xi)(\xi-1)
 \,  \re^{- {\ri \eta \xi p\p} }
 \int\limits_0^\infty\! d\l\,
 \exp\! \left[{-\frac{%\eta^2
 \p^2}{4\l}-\l\left[m^2 -(\xi-1)\xi p^2\right]} \right]
 \,,
\end{align}
where we used $\eta^2 =1$, and
by virtue of $p\tilde p =0$ we have $ - \inv{2}\sum\limits_{\eta=\pm 1}  \,  \re^{- {\ri \eta \xi p\p} } =-1$.
Carrying out the integration over $\lambda$ we end up with an integral over
a modified Bessel function $K_1(z)$ of second kind for expression \eqref{Jp}:
\begin{align}
\check J (p )  &=
\pi ^2 p^2 \int_0^1d\xi \ \frac{ (1-2 \xi )(\xi-1) \ K_1\left(\sqrt{\left(m^2-(\xi -1)
   \xi  p^2\right) \tilde{p}^2}\right)}{\sqrt{\frac{m^2- (\xi -1) \xi p^2}{\tilde{p}^2}}}
\,.
\end{align}

%===CONTENT ENDS HERE==========================================================

%\bibliographystyle{./../../custom1}
%\bibliography{./../../articles,./../../books,./../../wip}

%%%%%%%%%%%%%%%%%%%%%%%%%%%%%%%%%%%%%%%%%%%%%%%%%%%%%%%%%%%%%%%%%%%%%%%
% \newpage

\end{document}